\documentclass[a4,12pt]{article}
\usepackage{bm}
\usepackage{amsmath}
\usepackage{epsfig}
\newcommand{\bfvec}[1]{\hbox{\boldmath$#1$\unboldmath}}
\begin{document}
\begin{center}
{\bf \large QCD
susceptibilities and nuclear matter saturation in a chiral theory: inclusion of 
pion loops }\\[2ex]
G. Chanfray$^1$, M. Ericson$^{1,2}$\\ 
$^1$ IPN Lyon, Universit\'e Lyon 1 et 
 IN2P3-CNRS , F-69622 Villeurbanne Cedex\\
$^2$ Theory division, CERN, CH-12111 Geneva

\begin{abstract}
We derive the equation of state of symmetric nuclear matter in a relativistic theory with $\sigma$
and $\omega$ exchange. We take a chiral version of this model which insures all the chiral
constraints. Going beyond the mean field approach we introduce the effects of the pion
loops.  For the parameters of the model, in order to fix those linked to pion exchange, we exploit 
the most recent information  on the short-range part of the spin-isospin interaction. For 
those linked to the scalar  meson exchange we make  use of  an analysis of lattice results  on the 
nucleon mass evolution with the quark mass. With these inputs we are able reach a
correct description of the saturation properties. From the equation of state of symmetric nuclear matter 
we alsoderive the density dependence of the quark condensate and of the QCD susceptibilities.

\end{abstract}
Pacs{24.85.+p 11.30.Rd 12.40.Yx 13.75.Cs 21.30.-x} 
\end{center}

\section{Introduction}
                                                                       
Relativistic theories of nuclear matter are needed for the prediction of the
equation of state of nuclear matter at large densities, or of other quantities
such as the QCD quark condensate or the QCD susceptibilities. They find a natural framework 
in the relativistic $\sigma$ and $\omega$ exchange model \cite{SW86}. 
One important aspect is its chiral nature,  especially if it has  to be used also  for 
the investigation of QCD related quantities such as the quark condensate. The approach that we follow 
in this work for symmetric nuclear matter is based on the chiral version of the $\sigma$ and 
$\omega$ model that we have proposed  in  ref. \cite{CEG02}. 
Here the sigma model is formulated  in a non-linear version but with the presence of 
a chiral singlet scalar field $s$. The nuclear binding is provided by this chiral invariant
field and in this way  all chiral constraints are automatically satisfied. However it is  known that in 
such chiral theories attractive tadpole diagrams destroy saturation \cite{KM74,BT01}. Hence in order
to achieve saturation we complement our model, as in our previous work \cite{CE05},  by the
introduction of the scalar nucleonic response  first introduced in the quark-meson coupling
model, QMC \cite{G88}.  In  our previous investigation \cite{CE05},  we have applied this model in
the Hartree approximation to derive the equation of state. From this we have deduced the density evolution
of the quark condensate and of the QCD susceptibilities. This procedure insures the compatibility
with the saturation properties of nuclear matter. Our conclusions were that the deviations
of the condensate evolution with density from the linear behavior are mild and that there exists a 
convergence effect between the scalar and pseudoscalar susceptibilities, a conclusion 
already reached in ref. \cite{CE03,CEG03}.

The pion which does not contribute at the mean field level was ignored in this work.
However the exchange of the scalar $s$  field between nucleons is not the only source of 
attraction. Beside the exchange of a scalar meson, the Van Der Waals type of forces, {\it i.e.},   two-pion 
exchange with one or two intermediate Delta  excitations, also contribute to the middle range 
attraction of the NN potential and to the nuclear binding. Iterated  pion exchange is also  a
source of attraction. It is even envisaged in a chiral theory 
that it could describe the bulk of the nuclear binding \cite{KFW02} .
It is therefore natural to extend our previous descriptions restricted to the Hartree
scheme so as to incorporate pion loops, which is the aim of the present article. When pion
exchange forces are discussed one has also  to introduce the other components of the spin-isospin force,
namely the short-range contribution embodied in the Landau Migdal parameters. We will also introduce 
rho exchange, which acts in the spin transverse channel. We will show that the role
of both the short-range terms and the rho in the correlation energy is crucial. For what concerns the
values of the  Landau-Migdal parameters, years of experimental and theoretical investigations have
sharpened our knowledge of these quantities. We will exploit here the most recent results, for which a
survey is given in a recent article by Ichimura {\it et al.} \cite{ISW06}.
We feel timely to revisit certain problems  of nuclear physics and combine in these studies  
the experience accumulated in the pionic type of physics including chiral aspects and some results of 
QCD to have simultaneously a description of the nuclear binding and the evolution of chiral symmetry 
restoration, {\it i.e.}, of the quark scalar density, and of the QCD susceptibilities.

In order to achieve this we will incorporate in a consistent way, on top of this mean field, the pion 
loop contribution to the energy, namely the pion Fock term  and the pionic correlation energy,
taking into account short-range correlation effects, within a RPA ring approximation
\cite{CSN90}. 
In this way, by taking the successive derivatives of the grand potential
with respect to the pion mass, we will obtain the full contribution to the evolution of the chiral 
condensate and the scalar susceptibility.
In a recent work \cite{CDEM06}
we obtained the pionic piece of the scalar susceptibility by taking the
derivative with respect to the quark mass (in practice to the pion mass) of the pion 
scalar density $\langle\Phi^2\rangle$, which is a component of the condensate of the medium. 
We showed that the medium effects affecting the scalar
susceptibility are precisely the same that the ones governing the reshaping of the strength in 
the low invariant mass region in two-pion production experiments \cite{BO96,ST00,M02}. 
One purpose of the present paper is to provide a unified description of  two quantities,
{\it i.e.} the effect of the nuclear scalar field and of the pion loops in a way compatible 
with saturation properties of nuclear matter.  The pionic contributions  depend on the pionic 
polarization propagator {\it i.e.} the $p.h.$ (particle-hole) and $\Delta.h.$ (Delta-hole) Lindhardt 
functions. Although we start from a relativistic mean field theory we will evaluate these Lindhardt 
functions in a non-relativistic 
framework  for convenience since it is based on a well established phenomenology. 
It is interesting to point out that this approach where pion loops are
incorporated on top of background scalar and vector mean-fields has some similarities with 
the approach of reference \cite{FKVW06} based on a relativistic density functional.
There are however important differences that  will be discussed in section 4.

Our article is organized as follows. Section 2 is devoted to the loop energy. After a brief summary
of  the results obtained in the Hartree scheme  pion loops are introduced together with the other 
short-range components of the spin-isospin interaction. Section 3 discusses the QCD quantities, the
quark condensate and the scalar susceptibility. Finally in section 4 numerical results are given and 
the results discussed.

\section{Pion loop energy}
\subsection{Summary of the mean field results without pion loop}

We start with the mean field Hamiltonian used in ref. \cite{CE05} supplemented by the free pion
Hamiltonian~:
\begin{equation}
H_0=\int d{\bf r}\,\bar N\left(-i\vec\gamma\cdot\vec\nabla + M^*_N(s)+ g_\omega\gamma^0\omega_0\right)\,N
\,+\,V(s) -{1\over 2}\,m^2_\omega \omega_0^2\,+\,\sum_{{\bf q}, j}{\omega_q\over 2}\,
b^\dagger_{{\bf q} j} b_{{\bf q} j}
\label{HAM}
\end{equation}
where $s$ is the chiral invariant field associated with the 
radius $S=f_\pi\,+\,s$ of the chiral circle. $V(s)=V_0(s)\,-\, c S$ is the vacuum
potential which can be split into $V_0(s)=(\lambda/4)\big((f_\pi+s)^2-v^2\big)^2$, 
responsible for spontaneous chiral symmetry 
breaking, and the explicit symmetry breaking piece, $- c S$, where $c= f_\pi m_\pi^2$. 
As usual in relativistic mean field theories  we have added a coupling to an omega field $\omega_0$~.

At the mean field level the energy density for symmetric nuclear matteris given by~:
\begin{equation}
{E_0\over V}=\varepsilon_0=\int\,{4\,d^3 p\over (2\pi)^3} \,\Theta(p_F - p)\,E^*_p(\bar s)
\,+\,V(\bar s)\,+\,{g^2_\omega\over 2\, m_\omega^2}\,\rho^2
\end{equation}
and $E^*_p(\bar s)=\sqrt{p^2\,+\,M^{*2}_N(\bar s)}$ is the energy of an 
effective nucleon with the effective mass $M^*_N(\bar s)$. 
The effective mass is sensitive to the effect of the nucleon susceptibility, $\kappa_{NS}$,  
which embeds the influence of the internal nucleon structure~: 
\begin{equation}
M^*_N(\bar s)=M_N\,\left( 1+{\bar s\over f_{\pi }}\right)\,+{1\over 2}\,\kappa_{NS}\,
\bar s^2 .\label{EFFMASSN}
\end{equation}
The expectation value, $\bar S= f_\pi + \bar s$, of the $S$ field plays the role 
of a chiral order parameter. It is obtained by minimizing the energy density~:
$\partial\varepsilon/ \partial\bar s= g^*_S\,\rho_S\,+\,V'(\bar s)=0$ 
with the following expressions for the scalar density, $\rho_S$, 
and the scalar coupling constant $g^*_S$~:  
$$\rho_S=\int\,{4\,d^3 p\over (2\pi)^3} \,\Theta(p_F - p)\,{M^*_N\over E^*_p}
\quad\hbox{and}\quad g^*_S(\bar s)={\partial M^*_N\over\partial\bar s}={M_N\over f_{\pi}}
\,+\,\kappa_{NS}\, \bar s . $$
 The quantity $g_S=M_N/f_\pi$ is the scalar coupling constant 
of the model in the vacuum. Notice that the density dependence of $g^*_S$  arises entirely from 
the susceptibility term. Since the mean scalar field is negative and the sign of  $\kappa_{NS}$ 
is positive, $g^*_S$ is a decreasing function of the density. The in-medium sigma mass is obtained as 
the second derivative of the energy density with respect to the order parameter~:
$m^{*2}_\sigma =\partial^2 \varepsilon/\partial\bar s^2=V''(\bar s)\,+\,
\partial\left(g^*_S\,\rho_S\right)/ \partial \bar s$.

In  the  free pion part of the Hamiltonian of eq. \ref{HAM},  the operator $b^\dagger_{{\bf q} j}$ 
creates a free pion state with momentum $\bf q$ and isospin index $j$ and we we have 
$\omega_q=\sqrt{q^2+m^2_\pi}$ where $m_\pi$ is the pion mass. In fact in a more complete treatment 
including s-wave pion-nucleon coupling, according to ref. \cite{CEG02}, the pion mass should be
replaced by an effective mass $m^{*2}_\pi=m^2_\pi(1+\bar s/f\pi)$. As in ref. 
\cite{LBT06} we do not consider this  effect since  it is almost completely compensated by other 
contributions \cite{CEO03} and the pion mass remains stable in the nuclear medium \cite{DEE92}. 

 \subsection{Inclusion of the pion-nucleon coupling}

We now include the usual derivative  pion-nucleon interaction Hamiltonian~: 
\begin{equation}
H_{\pi NN}=-\,\int d{\bf r}\,{g_A\over f_\pi} \bar N \,\gamma^\mu\, \gamma_5\,
{\vec\tau\over 2}.\partial_\mu \vec\Phi\, N\simeq -\,\int d{\bf r}\,
{g_A\over 2\,f_\pi}\, N^\dagger\,
\bfvec{\sigma}.\bfvec{\nabla} \vec\Phi.\vec\tau\, N\label{LPINN}
\end{equation}
where the second form corresponds to the non relativistic limit of pure p-wave nature.
It can be rewritten in a standard second-quantized form as~: 
\begin{eqnarray}
H_{\pi NN}&=&\sum_{{\bf q},j}\,\left({1\over 2\,\omega_q\,V}\right)^{1/2}\,
\left(b_{{\bf q},j}\,+\,b^\dagger_{-{\bf q},j}\right)\,\,L_j({\bf q})\nonumber\\
L_j({\bf q})&=&
\sum_{\alpha,\beta}\,c^\dagger_\beta\,c_\alpha\,V_j^{\beta\alpha}({\bf q}),\quad
V_j^{\beta\alpha}({\bf q}, \omega)=-i
\,{g_A\over 2\,f_\pi}\, v({\bf q})\,
<N : \beta|\, \bfvec{\sigma}.{\bf q}\, \tau_j \, 
e^{i{\bf q}.{\bf R}}|N :\alpha\rangle 
\label{HINT}
\end{eqnarray}
where we have included phenomenologically a dipole $\pi NN$ form factor 
$v({\bf q})$ and $V$ is the volume of the normalization box.
Using  Green's function techniques, the expectation value on the nuclear
ground state $|0\rangle$ of this interacting Hamiltonian can be obtained according to \cite{CSN90}~:
\begin{equation}
\langle H_{\pi NN}\rangle=3\,V\,\int_{-\infty}^{+\infty} {i\,d\omega\over 2\pi)}\int{d{\bf q}\over (2\pi)^3}
\,{1\over \omega^2\,-\,\omega^2_q}\,\Pi_L(\omega, {\bf q}).\label{BASIC}
\end{equation}
Here $\Pi_L(\omega, {\bf q})$ is the full spin-isospin polarization propagator defined by~:
\begin{eqnarray}
\Pi_L(\omega, {\bf q})&=& \int_{-\infty}^{+\infty} d(t-t')\,e^{i\omega(t-t')}\,
{(-i)\over V}\,\langle 0| T\left(L_j^\dagger({\bf q}, t)\,,\,L_j({\bf q}, t')\right)|0\rangle
\nonumber\\
&=&{1\over V}\,\sum_n\,{|\langle n|\, L_j({\bf q})|0\rangle|^2\over \omega\,-\,E_n\,+\,i\eta}
\,-\,{|\langle n|\, L_j^\dagger({\bf q})|0\rangle|^2\over \omega\,+\,E_n\,-\,i\eta}\, .
\end{eqnarray}
We also include the Delta-hole states. In this case the operator 
$\bfvec{\sigma}.{\bf q}\, \tau_j$ is replaced by $(g_{\pi N\Delta}/g_{\pi NN})
\,\bf{S}.{\bf q}\, \,T_j$ where the $S_j$ ($T_j$) are the standard spin (isospin) transition 
operators  between spin (isospin) $1/2$ and $3/2$ states. For the ratio of coupling 
constants, we take the  quark model  value~:  
$R_{N\Delta}=(g_{\pi N\Delta}/g_{\pi NN})=\sqrt{72/25}$.

 \subsection{Inclusion of short-range correlation and rho meson exchange }
The expression (\ref{BASIC}) written above shows that the interaction energy  of the
pion cloud with matter   is  obtained by integrating the product of the full longitudinal 
spin-isospin polarization propagator ({\it i.e.} the nuclear response to a pion-like excitation)
with the pion exchange potential. However, due to the correlated nature of the medium, the contact piece  
of the pion exchange potential should be treated with caution. In the part which concerns medium 
effects, this piece  should  be removed. This is   achieved as usual by introducing   in the spin-isospin 
channel a short-range two-body potential governed by  Landau-Migdal $g'$ parameters.
The $Nh-Nh$ piece of the corresponding Hamiltonian writes~:
\begin{eqnarray}
H_{sr}&=&{1\over 2}\,\sum_{\alpha\beta\gamma\delta} c^\dagger_\alpha c^\dagger_\beta c_
\delta c_\gamma\,
\langle\alpha\beta |V_{sr}(1,2)|\gamma\delta\rangle\nonumber\\
V_{sr}(1,2)&=&\int{d{\bf q}\over (2\pi)^3}
\,\left({g_A\over 2\,f_\pi}\right)^2 v^2({\bf q})
\,g'_{NN}\,\bfvec{\sigma}(1).\bfvec{\sigma}(2) 
\, \vec\tau(1).\vec\tau(2)\,
e^{-i{\bf q}.({\bf R}_1-{\bf R}_2)}\, .
\end{eqnarray}
It is  straightforward to extend this expression in  order to include the $Nh-\Delta h$ 
and the $\Delta h-\Delta h$ pieces. For this we replace $g'_{NN}$ by $R_{N\Delta}\, g'_{N\Delta}$
and $R_{N\Delta}^2\, g'_{\Delta\Delta}$ where $R_{N\Delta}$ defined previously is the
coupling constant ratio. In the numerical calculation we will use the following  set of values 
suggested in ref. \cite{ISW06}, which takes into account the most recent data analysis~: 
$g'_{NN}=0.7,\,g'_{N\Delta}=0.3,\,g'_{\Delta\Delta}=0.5$, with clear deviations from universality as 
advocated in ref. \cite{ISW06}.
The spin-spin operator can be decomposed into a longitudinal piece and a transverse one.
For this reason we also introduce the full transverse spin-isospin polarization propagator~: 
\begin{eqnarray}
\Pi_T(\omega, {\bf q})&=& {1\over 2}\,\int_{-\infty}^{+\infty} d(t-t')\,e^{i\omega(t-t')}\,
{(-i)\over V}\,\langle 0| T\left({\bf T}_j^\dagger({\bf q}, t)\,,\,{\bf T}_j({\bf q}, t')\right)
|0\rangle\nonumber\\
&=&{1\over 2 V}\,\sum_n\,{|\langle n|\,{\bf T}_j({\bf q})|0\rangle|^2\over \omega\,-\,E_n\,+\,i\eta}\,-\,
{|\langle n|\,{\bf T}_j^\dagger({\bf q})|0\rangle|^2\over \omega\,+\,E_n\,-\,i\eta}
\end{eqnarray} 
where ${\bf T}_j$ is obtained from $L_j$ by replacing $\bfvec{\sigma}.{\bf q}$ by 
$\bfvec{\sigma}\times{\bf q}$. The expectation value of the short-range potential is ~: 
\begin{equation}
\langle H_{sr}\rangle={3\over 2}\,V\,\int_{-\infty}^{+\infty} {i\,d\omega\over 2\pi)}\int{d{\bf q}\over 
(2\pi)^3}
\,{g'}\, \big[\,\left(\Pi_L(\omega, {\bf q})\,+\,2\,\Pi_T(\omega, {\bf q})\right)\big]\,-\,
{1\over 2}\,\sum_{\alpha h}\,\langle\alpha h |V_{sr}(1,2)|h \alpha\rangle
\end{equation}
where the second term is introduced in order to remove the contribution for free nucleons. In the 
following, when this second term term is not explicitely written,  this  substraction is done 
implicitely. The first  term ${g'}\,\left(\Pi_L(\omega, {\bf q})\,+\,2\,\Pi_T(\omega, {\bf
q})\right)$ is written in a schematic form. The precise form involving the three
Landau-Migdal  parameters, $g'_{NN},g'_{N\Delta}, g'_{\Delta\Delta}$, will be given below
in subsection {\bf 2.4}.

Finally, so as to obtain a more realistic description of the transverse spin-isospin channel we 
also introduce, beside the short-range piece,  the $\rho$ meson through  a $\rho NN$ Hamiltonian. 
In the non relativistic limit it has a structure similar to the $\pi NN$ one, 
once the $\bfvec{\sigma}.{\bf q}$ coupling is replaced by a $\bfvec{\sigma}\times{\bf q}$
coupling. Its expectation value is~: 
\begin{equation}
\langle H_{\rho NN}\rangle=3\,V\,\int_{-\infty}^{+\infty} {i\,d\omega\over 2\pi)}\int{d{\bf q}\over 
(2\pi)^3}\,{2\,C_\rho\over \omega^2\,-\,\Omega^2_q}\,\Pi_T(\omega, {\bf q}).
\end{equation}
with $\Omega_q=\sqrt{q^2\,+\,m_\rho^2}$. We take $C_\rho=2$ which corresponds to the
strong rho coupling \cite{HP75}.
 \subsection{Calculation of the pion loop energy }
In order to calculate the pion loop energy we use the well-known charging formula method.
For this purpose we introduce an auxiliary Hamiltonian which depends on one strength 
parameter $\lambda$~:
\begin{equation} 
H(\lambda)=H_0\,+\,\lambda\,(H_{\pi NN}\,+\,H_{\rho NN})\,+\,\lambda^2\,H_{sr}\, .
\end{equation}
$H(\lambda)$ coincides with the original Hamiltonian for $\lambda=1$. The charging formula
allows the evaluation of the modification of the ground state energy with respect 
to the ground state energy $E_0$ of the $H_0$ Hamiltonian~:
\begin{eqnarray}
E_{loop}=E\,-\,E_0&\equiv& V\,\varepsilon_{loop}={3\over 2}\,V\,\int_{-\infty}^{+\infty} 
{i\,d\omega\over 2\pi)}\int{d{\bf q}\over 
(2\pi)^3}\,\int_0^1{d\lambda\over\lambda}\nonumber\\
& &\left( \big[V_L(\omega, {\bf q})\,\Pi_L(\omega, {\bf q}; \lambda)\big]\,+\,2\,
\big[V_T(\omega, {\bf q})\,\Pi_T(\omega, {\bf q}; \lambda)\big]\right)~.\label{ELOOP}
\end{eqnarray}
The dependence  on $\lambda$ of  the full polarization propagators can be obtained by
systematically multiplying the coupling $g_A/ f_\pi$ by $\lambda$. For instance 
there appears  a net 
$\lambda^2$ factor in front of each first order Lindhardt function appearing in the RPA expression of the
full polarization propagators.The quantities 
$V_{L,T}(\omega, {\bf q})$ are  the full energy dependent residual interactions in 
the longitudinal and transverse spin-isospin channels. 
In the $Nh-Nh$ sector,  they write~:
\begin{equation}
V_{LNN}(\omega, {\bf q})=g'_{NN}\,+\,{q^2\over \omega^2\,-\,\omega_q^2},\qquad
V_{TNN}(\omega, {\bf q})=g'_{NN}\,+\,C_\rho\,{q^2\over \omega^2\,-\,\Omega_q^2},\qquad
\end{equation}
with identical expressions in the  $Nh-\Delta h$ 
and the $\Delta h-\Delta h$ channels once the relevant g' parameters have been
incorporated. Again the expression (\ref{ELOOP}) written above is schematic and the 
full formula calculated in
the RPA ring approximation is given just below in eqs. \ref{EPSLONG},\ref{EPSTRAN}. 
The $\lambda$ integration can 
be done  analytically to obtain the loop energy density $\varepsilon_{loop}$~:
\begin{equation} 
\varepsilon_{loop}=\varepsilon_L\,+\,\varepsilon_T \qquad \hbox{with}
\end{equation}
\begin{equation}  
\varepsilon_L={3\over 2}\int {id\omega d{\bf q}\over 
(2\pi)^4}\big[-ln\big(1-V_{LNN}\Pi^0_N
-V_{LN\Delta}\Pi^0_\Delta-(V^2_{LN\Delta}-V_{LNN}\,V_{L\Delta\Delta})
\Pi^0_N\Pi^0_\Delta\big)\big]\label{EPSLONG}
\end{equation}
\begin{equation}
\varepsilon_T={3}\int{id\omega d{\bf q}\over 
(2\pi)^4}\big[-ln\big(1-V_{TNN}\Pi^0_N
-V_{TN\Delta}\Pi^0_\Delta-(V^2_{TN\Delta}-V_{TNN}\,V_{T\Delta\Delta})
\Pi^0_N\Pi^0_\Delta\big)\big]\label{EPSTRAN}
\end{equation}
$\Pi^0_{N}$ is the $N.h.$ Lindhardt function which is calculated non relativistically.
With the single particle energy, $\varepsilon_p=p^2/2 M^*$ where $M^*$ is the $s$ dependent
nucleon effective mass calculated at the mean-field level,  $\Pi^0_{N}$ has the familiar
form~:
 \begin{eqnarray}
\Pi^0_{N} (\omega, {\bf q})= 4\, \left({g_A\over 2 f_\pi}\right)^2\, 
v^2({\bf q})\,
\int {d{\bf p}\over (2 \pi)^3}& &
\bigg({\Theta ( p_F-p)\,\Theta (|{\bf p}+{\bf q}|
-p_F)\over \omega\,-\,\epsilon_{N{\bf p}{\bf q}}\,+\,i\eta}
\nonumber\\
& &-\,{\Theta (p_F-p)\,\Theta (|{\bf p}+{\bf q}|
-p_F)\over \omega\,+\,\epsilon_{N{\bf p}{\bf q}}-\,i\eta}\bigg)\, .
\label{PI0N}\end{eqnarray}
$\epsilon_{N{\bf p}{\bf q}}=\epsilon_{{\bf p}+{\bf q}}\,-\,\epsilon_{\bf p}$ is the energy 
of the p.h. excitation. 
The corresponding $\Delta.h.$ Lindhardt function, $\Pi^0_{\Delta}$ is also calculated non 
relativistically. Introducing also an effective $\Delta$ mass, we make the natural assumption  that the 
in-medium shift of the nucleon and delta masses in presence of the nuclear scalar field are 
identical. The $\Delta.h.$ energy is thus~:
$\epsilon_{\Delta{\bf p}{\bf q}}=\epsilon_{\Delta,{\bf p}+{\bf q}}\,-\,\epsilon_{\bf p}$
with $\epsilon_{\Delta,{\bf p}}=M_\Delta -M_N + p^2/2 M_\Delta^*$ and
$\Pi^0_{\Delta}$ writes~:
\begin{equation}
\Pi^0_{\Delta} (\omega, {\bf q})= {16\over 9}\, \left({g_A\over 2 f_\pi}\,R_{N\Delta}\right)^2
\, v^2({\bf k})\,
\int {d{\bf p}\over (2 \pi)^3}\,\Theta (p_F-p)
\bigg({1\over \omega\,-\,\epsilon_{\Delta{\bf p}{\bf q}}\,+\,i\eta}
-\, {1\over 
\omega\,+\,\epsilon_{\Delta{\bf p}{\bf q}}}\bigg)_, .
\label{PI0D}
\end{equation}
This loop energy can be split in a piece of mean-field nature (containing the Fock term) 
and a genuine correlation energy. For instance we have for the longitudinal piece~:
\begin{eqnarray}
\varepsilon_L &=& \varepsilon^{MF}_L \,+\,\varepsilon^{Corr}_L \nonumber\\
\varepsilon^{MF}_L &=&{3\over 2}\int {id\omega d{\bf q}\over 
(2\pi)^4}\,\big[V_{LNN}\Pi^0_N + V_{LN\Delta}\Pi^0_\Delta\big]\nonumber\\
\varepsilon^{Corr}_L&=&{3\over 2}\int {id\omega d{\bf q}\over 
(2\pi)^4}\big[-ln\big(1-V_{LNN}\Pi^0_N
-V_{LN\Delta}\Pi^0_\Delta-(V^2_{LN\Delta}-V_{LNN}\,V_{L\Delta\Delta})
\Pi^0_N\Pi^0_\Delta\big)\nonumber\\
& &- V_{LNN}\Pi^0_N-V_{LN\Delta}\Pi^0_\Delta\big]\, .\label{EL}
\end{eqnarray}
By performing the energy integration with a Wick rotation, $\varepsilon_L^{MF}$ 
can be expressed as follows~:
\begin{equation}
\varepsilon^{MF}_L=
\int {4 d{\bf p}\over (2 \pi)^3}\Theta ( p_F-p)\,\Sigma_\pi({\bf p})\quad+\quad
\varepsilon^{Fock}_L\, . 
\end{equation}
$\Sigma_\pi({\bf p})$ is the pionic   contribution to the nucleon self-energy~:
\begin{equation}
\Sigma_\pi({\bf p})=-{3\over 2}\left({g_A\over 2 f_\pi}\right)^2\int{d{\bf q}\over 
(2\pi)^3} {\bf q}^2 \,v^2({\bf q})\left({1\over \omega_q}{1\over \omega_q+\epsilon_{N {\bf p}{\bf q}}}
+{4 R_{N\Delta}^2\over 9}{1\over \omega_q}{1\over \omega_q+\epsilon_{\Delta{\bf p}{\bf
q}}}\right)\, .
\end{equation}
In principle medium effects are present in this quantity since the $N.h.$ and $\Delta.h.$ energies 
depend on the effective nucleon and delta masses but in practice  we ignore these effects. 
The  momentum dependence of the nucleon self-energy is linked to the treatment of  
the vacuum nucleon which is not the purpose of this paper. Hence  we ignore it. At zero momentum the 
self-energy $\Sigma_\pi\equiv \Sigma_\pi({\bf p}=0)$ is similar to the   model calculation one  
used to fit lattice data by Thomas et al \cite{TGLY04}, although we do not make a static approximation.
In our calculation we will choose a dipole form factor, with a cutoff $\Lambda=0.98\,GeV$, such that 
the resulting  
contribution to the free nucleon sigma  term, $\sigma^{(\pi)}_N$, the  expression  of which being given in 
subsection {\bf 3.1}, is $\sigma^{(\pi)}_N=21.5\, MeV$, in agreement with previous determinations
\cite{JTC92,BM92}. The corresponding  pion cloud self-energy is $\Sigma_\pi=- 420\, MeV$.

The Fock term is~: 
\begin{equation}
\varepsilon^{Fock}_L=
-{3\over 2}\left({g_A\over 2 f_\pi}\right)^2\int{d{\bf q}\over 
(2\pi)^3}  v^2({\bf q}) \int {4 d{\bf p}\over (2 \pi)^3}\Theta(p_F-p)  
\Theta(p_F-|{\bf p}+{\bf q}|)
\left(g'_{NN}-{{\bf q}^2\over\omega_q (\omega_q+\epsilon_{N{\bf p}{\bf q}})}\right).
\end{equation}
Notice that this Fock term includes a retardation effect through the presence of the
particle-hole energy $\epsilon_{N{\bf p}{\bf q}}$. In practice the effect is numerically small
and we take it as its average Fermi sea value~: $\epsilon_{N{\bf p}{\bf q}}=\epsilon_{Nq}=q^2/2 M_N$, 
ignoring the in-medium modification of the nucleon mass. 
In that way the Fock energy  does not depend on
$\bar s$, a feature which simplifies the calculation of the scalar susceptibility. 
Also notice that, at variance with other works, see e.g. \cite{BMVN87}, we keep the form factor in the 
Fock term.  The Fock term relative to the transverse channel is obtained in a similar way~:
\begin{equation}
\varepsilon^{Fock}_T=
-{3}\left({g_A\over 2 f_\pi}\right)^2\int{d{\bf q}\over 
(2\pi)^3}  v^2({\bf q}) \int {4 d{\bf p}\over (2 \pi)^3}\Theta(p_F-p)  
\Theta(p_F-|{\bf p}+{\bf q}|)
\left(g'_{NN}-{C_\rho\, {\bf q}^2\over\Omega_q (\Omega_q+\epsilon_{N{\bf p}{\bf
q}})}\right)
\end{equation}
Finally the correlation energy is also calculated using a Wick rotation for the 
$\omega$ energy integration. The numerical results will be given in section {\bf 4}.

\section{The chiral condensate and the scalar susceptibility}
\subsection{ Direct calculation} 
In this first approach the in-medium quark condensate is evaluated directly from the expectation value
of chiral symmetry breaking piece of the Hamiltonian~:
\begin{equation}
\langle H_{\chi SB}\rangle\equiv V\,2\,m\,\langle\bar q q\rangle\simeq V\,\left\langle{1\over 2}\,m^2_\pi\,\Phi^2
\,-\, c\,S\right\rangle\quad\to\quad\langle\bar q q\rangle\simeq\langle\bar q q\rangle_{vac}\left(1\,-\,
{\langle\Phi^2\rangle\over 2\,f_\pi^2}\,+\,
{\bar s\over f_\pi}\right)
\end{equation}
where  the GOR relation has been used to obtain the second equation. The pion scalar density
can be directly calculated from the in-medium pion propagator \cite{CD99,CDEM06}~:
\begin{equation}
\langle\Phi^2\rangle=3\,\int{id\omega d{\bf q}\over 
(2\pi)^4}\,\left[D_\pi - D_{0\pi}\right](\omega,{\bf q})=
3\,\int{id\omega d{\bf q}\over 
(2\pi)^4}\,\left[D_\pi  D_{0\pi}(\omega,{\bf q})\,{\bf q}^2\,\tilde\Pi^0(\omega,{\bf q})\right]\,.
\label{PHI2}\end{equation}
$D_\pi=(\omega^2-\omega_q^2-{\bf q}^2\tilde\Pi^0)^{-1}$ is the full pion propagator,
${\bf q}^2\tilde\Pi^0$ is the irreducible pion self-energy with~:
\begin{equation}
\tilde\Pi^0={\Pi^0_N +\Pi^0_\Delta + (2g'_{N\Delta}-  g'_{NN} - g'_{\Delta\Delta} )  
\Pi^0_N \Pi^0_\Delta\over
1- \big(g'_{NN}\Pi^0_N + g'_{\Delta\Delta}\Pi^0_\Delta +
(g'^2_{N\Delta}-  g'_{NN}  g'_{\Delta\Delta} )\Pi^0_N \Pi^0_\Delta\big)}\, .
\end{equation}
It is possible to show that the formal expression of the pion scalar density can be
 obtained as the
derivative with respect to $m_\pi^2$ of the pionic energy density $\varepsilon_L$
 given by eq. \ref{EPSLONG}~:
\begin{equation}
{\langle\Phi^2\rangle\over 2}={\partial\varepsilon_L\over\partial m_\pi^2}\quad\hbox{or
equivalently}
\quad
- \langle\bar q q\rangle_{vac}{\langle\Phi^2\rangle\over 2\,f_\pi^2}=
{1\over 2}\,{\partial\varepsilon_L\over\partial m}
\end{equation}
which is nothing but the Feynman-Hellman theorem applied to the pion loop contribution.
As in the previous section we decompose $\varepsilon_L$ into a free nucleon contribution,
a Fock contribution and a correlation piece, with a similar decomposition for ${\langle\Phi^2\rangle}$ ~:
\begin{equation}
{\langle\Phi^2\rangle\over 2}=\rho\,{\partial\Sigma_\pi\over\partial m_\pi^2} +
{\partial\varepsilon_L^{Fock}\over\partial m_\pi^2} +{\partial\varepsilon_L^{Corr}\over\partial
m_\pi^2} \equiv 
\rho\,{\sigma_N^{(\pi)}\over m^2_\pi} + {\langle\Phi^2\rangle^{Fock}\over 2} +
{\langle\Phi^2\rangle^{Corr}\over 2}\, .
\end{equation}
The pionic contribution to the pion-nucleon sigma term is~:
\begin{eqnarray}
\sigma_N^{(\pi)} ={3\over 2}\,\left({g_A\over 2 f_\pi}\right)^2\,m_\pi^2\,\int{d{\bf q}\over 
(2\pi)^3}& & {{\bf q}^2 \,v^2({\bf q})\over 2\omega^2_q} 
\bigg[{1\over \omega_q}\,{1\over \omega_q+\epsilon_q}\,+\,{1\over (\omega_q+\epsilon_q)^2}
\nonumber\\
& &+{4\, R_{N\Delta}^2\over 9}\left({1\over \omega_q}\,{1\over \omega_q+\epsilon_{\Delta q}}
\,+\,{1\over (\omega_q+\epsilon_{\Delta q})^2} \right)\bigg]\, .
\label{PISIGMA}
\end{eqnarray}
The Fock term contribution has the following explicit form~:
\begin{eqnarray}
{\langle\Phi^2\rangle^{Fock}\over 2}=
-{3\over 2}\left({g_A\over 2 f_\pi}\right)^2 \,\rho\,
\int_0^{2 p_F}{d{\bf q}\over (2\pi)^3} 
{{\bf q}^2 \,v^2({\bf q})\over 2\omega^2_q(\omega_q+\epsilon_q)}& & 
\left(1 -{3 q\over 4 p_F} +{q^3\over 16 p_F^3}\right)\nonumber\\  
& &\left({1\over \omega_q}\,+\,{1\over (\omega_q+\epsilon_q)} \right).\label{PHIFOCK}
\end{eqnarray}
Finally the correlation piece is obtained by removing the leading order piece in 
eq. \ref{PHI2}~:
\begin{equation}
{\langle\Phi^2\rangle^{Corr}\over 2}= {3\over 2}\,\int{id\omega d{\bf q}\over 
(2\pi)^4}\,\,{\bf q}^2\,\left[D_\pi  D_{0\pi}\,\tilde\Pi^0
- D^2_{0\pi}\,\,\Pi^0\right](\omega,{\bf q}).\label{PHICORR}
\end{equation}
In the next subsection we will evaluate the condensate from the equation of state (grand potential)
and the Feynman-Hellman theorem. It leads essentially to the same result with 
a  modification coming from the fact that we will have to face the problem 
of the Lorentz nature of the in-medium effects of the pion loop energy.

\subsection{ The quark condensate from the equation of state} 
The pion cloud contribution has been calculated within the standard non relativistic 
framework. One reason is the  basis of the pion-nucleus 
phenomenology. However some care should be taken when this pion cloud is included on top of the 
relativistic mean field used to describe the coupling of the scalar field to nucleons.
For free nucleons the pionic self-energy $\Sigma_\pi$ has to be 
of   scalar nature and thus it adds  to the bare nucleon mass: $M_N=M_0+\Sigma_\pi$.
For what concerns the in-medium contribution  to the energy we do not know if it is
of scalar or vector nature or a mixture of both. Here we  assume its vector nature 
and  add it to the energy density.  This assumption affects contributions arising from the 
pionic interaction between nucleons where it amounts to distinguishing between scalar and vector 
densities. In practice the effect of this difference is numerically quite  small. For the free nucleon 
part instead this distinction has a non negligible effect. The energy density thus writes~:
\begin{equation}
{E\over V}=\varepsilon=\int\,{4\,d^3 p\over (2\pi)^3} \,\Theta(p_F - p)\,E^*_p(\bar s)
\,+\,V(\bar s)\,+\,{g^2_\omega\over 2\, m_\omega^2}\,\rho^2\,+\,\varepsilon^{FC}
\end{equation}
where $\varepsilon^{FC}=\varepsilon_L^{Fock}\,+\,\varepsilon_T^{Fock}\,+\,
\varepsilon_L^{Corr}\,+\, \varepsilon_T^{Corr}$ summarizes the loop energy density
from the Fock term and correlations (pion + rho + short-range) which can be decomposed into a
longitudinal piece (pion + short-range) and a transverse piece (rho + short-range). 
The mean field $\bar s$ is obtained by
minimizing the energy density before the incorporation of the loop energy or, said
differently, ignoring the $\bar s$ dependance of $\varepsilon^{FC}$. This is actually true for the
Fock term  and this approximation has a small numerical incidence on 
the evolution of the condensate with inclusion of the correlation energy. 
It also follows from the independence on $\bar s$ of $\epsilon^{FC}$ that the in-medium 
sigma mass remains the same as in the mean-field case~: 
\begin{equation}
m^{*2}_\sigma ={\partial^2 \varepsilon\over\partial\bar s^2}=V''(\bar s)\,+\,\kappa_{NS}\,
\rho_S\,+\,g^*_S\,
{\partial \rho^*_S\over\partial \bar s}\,.\label{MSIGMA}
\end{equation}
As discussed in ref. \cite{CE05} the scalar susceptibility term (second term in $\kappa_{NS}$ 
on the r.h.s of eq. \ref{MSIGMA})  counterbalances the effect of the first term, the in-medium
chiral dropping of the sigma mass. The last term actually corresponds to the nuclear response associated 
with $N \bar N$ excitation. In practice it is small and it can be omitted \cite{CEG03}.

We will derive the in-medium chiral condensate and the QCD scalar susceptibility from
the equation of state since 
they are  related to the first and second derivatives  of the grand
potential with respect to the quark mass $m$ at constant chemical potential $\mu$. 
The baryonic chemical potential is obtained as~:
\begin{equation}
\mu={\partial\varepsilon\over \partial\rho}= E^*_F\,+\,{g^2_\omega \over m_\omega^2}
\,\rho\,+\,{\partial\varepsilon^{FC}\over\partial\rho}
\qquad\hbox{with}\qquad E^*_F=\sqrt{p_F^2\,+\,M^{*2}_N(\bar s)} .
\end{equation}
In order to obtain this result we have taken into account only the explicit density  dependence 
on the density $\rho$, {\it i.e.}, keeping $\bar s$ constant since $\partial \varepsilon/\partial\bar s=0$.
One deduces that the baryonic density is controlled by the chemical
potential according to~: 
\begin{equation}
\rho=\int\,{4\,d^3 p\over (2\pi)^3} \,
\Theta\left(\mu\,-\,E^*_p\,-\,{g^2_\omega\over m_\omega^2}\,\rho
\,-\,{\partial\varepsilon^{FC}\over\partial\rho}\right)\label{RHOB}
\end{equation}
while the scalar density writes~:
\begin{equation}
\rho_S=\int\,{4\,d^3 p\over (2\pi)^3}\, {M^*_N\over E^*_p}\,
\Theta\left(\mu\,-\,E^*_p\,-\,{g^2_\omega \over m_\omega^2}\,\rho
\,-\,{\partial\varepsilon^{FC}\over\partial\rho}\right).
\end{equation}
The grand potential, which is obtained through a Legendre transform, can be 
written in the following form~:
\begin{eqnarray}
\omega(\mu)&=&\varepsilon\,-\,\mu\,\rho\nonumber\\
&=&\int\,{4\,d^3 p\over (2\pi)^3} \left(E^*_p\,+\,{g^2_\omega \over
m_\omega^2}\,\rho\,+\,{\partial\varepsilon^{FC}\over\partial\rho}
-\,\mu\right)\,\Theta\left(\mu\,-\,E^*_p\,-\,{g^2_\omega \over
m_\omega^2}\,\rho
\,-\,{\partial\varepsilon^{FC}\over\partial\rho}\right)\nonumber\\
& &+\,V(s)\,-\,{g^2_\omega\over 2\, m_\omega^2}\,\rho^2
\,+\,\varepsilon^{FC}\,-\,\rho\,{\partial\varepsilon^{FC}\over\partial\rho}.
\end{eqnarray}
For the derivation of  the condensate and of the susceptibility we have to specify the symmetry 
breaking parameter,  which in QCD is the quark mass. In the context of this model it is the quantity 
$c=f_\pi m_\pi^2$ which enters the symmetry breaking piece of the potential.  In the application 
of the Feynman-Hellman theorem we use the explicit expression of 
$\partial c/\partial m$ given by the model to leading order in the quark mass $m$, 
{\it i.e.},  $\partial c/\partial m=-2 \langle\bar q q\rangle_{vac}/ f_\pi$.
As previously,  in the calculation of the derivative we only keep the explicit dependence on $c$~:
\begin{eqnarray}
\langle\bar q q\rangle={1\over 2} \left({\partial \omega\over\partial m}\right)_\mu =
{1\over 2} {\partial c\over\partial m}\left({\partial \omega\over\partial c}\right)_\mu 
&\simeq &- {\langle\bar q q\rangle_{vac}\over
f_\pi}\,\left(-\bar S\,+\rho_S\,\left({\partial M_N\over\partial c}\right)_{\bar S}
\,+\,{\partial\varepsilon_L^{FC}\over\partial c}\right)\nonumber\\
&\simeq &\langle\bar q q\rangle_{vac}
\left({\bar S\over f_\pi}\,-\,{m_\pi^2\over c}{\partial \Sigma_\pi\over\partial m_\pi^2}
\,-\,{m_\pi^2\over c}{\partial\varepsilon_L^{FC}\over\partial m_\pi^2}\right)\nonumber\\
&=& \langle\bar q q\rangle_{vac}\left(1\,+\,{\bar s\over f_\pi}\,-
\,{\langle\Phi^2\rangle\over 2 f^2_\pi}\right)\,.
 \label{CONDENSAT}
\end{eqnarray}
In the last line we have grouped the pionic contribution in a pion scalar density given
by~:
\begin{equation}
{\langle\Phi^2\rangle\over 2} =\rho_S\,{\sigma_N^{(\pi)}\over m_\pi^2} + {\langle\Phi^2\rangle^{Fock}\over 2} +
{\langle\Phi^2\rangle^{Corr}\over 2}\,.
\end{equation}
Notice that the pion scalar density deviates from the fully non relativistic one 
of eq. \ref{PHI2} since the leading order term behaves like $\rho_S$ in place of $\rho$.
To leading order in density we recover, as expected, the well-known result~:
\begin{equation}
\langle\bar q q\rangle=\langle\bar q q\rangle_{vac}\left(1\,-\,{ \sigma_N\,\rho\over
f_\pi^2\,m_\pi^2}\right),\quad \sigma_N=\sigma_N^{(\pi)}\,+\,\sigma_N^{(\sigma)},\quad
\sigma_N^{(\sigma)}=f_\pi\,g_S\,{m_\pi^2\over m^2_\sigma}\,.
\end{equation}
\subsection{The scalar susceptibility}
The in-medium scalar susceptibility is obtained  as a derivative at fixed $\mu$ of the chiral condensate
given in eq. \ref{CONDENSAT}. The dependence of the pion scalar density on the symmetry
breaking parameter ($c$ or $m_\pi$) receives two contributions~: one comes from the explicit
dependence on c and the other from the implicit dependence since, at fixed $\mu$, the density $\rho$
depends on $c$ according to eq. \ref{RHOB}.  Hence the scalar susceptibility takes the form~:
\begin{eqnarray}
\chi_S&=&\left({\partial\langle\bar q q\rangle\over\partial m}\right)_\mu=
-{1\over 2}\left({\partial c\over\partial m}\right)^2\,\left[\left({\partial\bar S \over\partial c}
\right)_\mu\,-\,{1\over 2 f_\pi}\,\left({\partial \langle\Phi^2\rangle\over\partial
c}\right)_\mu\right]\nonumber\\
&\simeq &-2\,{\langle\bar q q\rangle_{vac}^2\over f_\pi^2}\,
\left[\left({\partial\bar S \over\partial c}\right)_\mu\,
-\,{1\over 2 f_\pi}\,\left({\partial \langle\Phi^2\rangle\over\partial\rho}\right)_\mu 
\left({\partial \rho\over \partial c}\right)_\mu\,-\,{m_\pi^2\over 2 f_\pi c}
\,\left({\partial \langle\Phi^2\rangle\over\partial m_\pi^2}\right)\right]\nonumber\\ 
&\equiv &\chi_S^{nuclear}\,+\,\chi_S^{pion loop}.\label{CHIS}
\end{eqnarray}
We have divided $\chi_S$ in two components. The first one which is the sum of the first two terms 
contains the coupling of the scalar quark  density fluctuations to the nuclear $p.h.$ excitations. 
For this  reason we denote it $\chi_S^{nuclear}$~:

\begin{equation}
\chi_S^{nuclear}=-2\,{\langle\bar q q\rangle_{vac}^2\over f_\pi^2}\,
\left[\left({\partial\bar S \over\partial c}\right)_\mu\,
-\,{1\over 2 f_\pi}\,\left({\partial \langle\Phi^2\rangle\over\partial\rho}\right)_\mu\,
\left({\partial \rho\over \partial c}\right)_\mu\right]\,. 
\end{equation}

The second one, $\chi_S^{pion loop},$  of purely pionic nature, comes  from the explicit dependence 
on $m_\pi$  of the pion scalar density $\langle\Phi^2\rangle$. The latter can be written as~:
\begin{equation}
\chi_S^{pion loop}=
2\,{\langle\bar q q\rangle_{vac}^2\over f^2_\pi}\,{1\over 2 f^2_\pi }
\,{\partial \langle\Phi^2\rangle\over\partial m_\pi^2}=
\rho_S\,\chi_{NS}^{(\pi)}\,+\,\chi_S^{pion-Fock}\,+\,\chi_S^{pion-Corr}\,.\label{CHIPI}
\end{equation}
The pionic contribution to the nucleon scalar susceptibility $\chi_{NS}^{(\pi)}$
is obtained from the derivitative of the pion cloud contribution to the sigma commutator
(eq. \ref{PISIGMA}):
\begin{eqnarray}
\chi_{NS}^{(\pi)} =2\,{\langle\bar q q\rangle_{vac}^2\over f^4_\pi}& & 
{3\over 2}\,\left({g_A\over 2 f_\pi}\right)^2\int{d{\bf q}\over 
(2\pi)^3} {{\bf q}^2 \,v^2({\bf q})\over 2\omega^3_q}
\bigg[{3\over \omega^2_q \,(\omega_q+\epsilon_q)}\,+\,
{3\over \omega_q \,(\omega_q+\epsilon_q)^2}
\,+\,{2\over(\omega_q+\epsilon_q)^3}
\nonumber\\
& &+{4\, R_{N\Delta}^2\over 9}\left({3\over \omega^2_q\,(\omega_q+\epsilon_{\Delta q})}
\,+\, {3\over \omega_q\,(\omega_q+\epsilon_{\Delta q})^2}
\,+\, {2\over (\omega_q+\epsilon_{\Delta q})^3}\right)\bigg]\, .
\label{CHINS}
\end{eqnarray}
The Fock term contribution is obtained from eq. \ref{PHIFOCK}~:
\begin{eqnarray}
\chi_S^{pion-Fock}=- 2\,{\langle\bar q q\rangle_{vac}^2\over f^4_\pi}
\,{3\over 2}\left({g_A\over 2 f_\pi}\right)^2& &\rho\,
\int_0^{2 p_F}{d{\bf q}\over 
(2\pi)^3} {{\bf q}^2 \,v^2({\bf q})\over 2\omega^3_q (\omega_q+\epsilon_q)}
\left(1 -{3 q\over 4 p_F} +{q^3\over 16 p_F^3}\right) \nonumber\\ 
& &\left( {3\over \omega^2_q \,(\omega_q+\epsilon_q)}\,+\,
{3\over \omega_q \,(\omega_q+\epsilon_q)^2}
\,+\,{2\over(\omega_q+\epsilon_q)^3}\right)\,.\label{CHIFOCK}
\end{eqnarray}
Finally the correlation piece is obtained from eq. \ref{PHICORR}~: 
\begin{equation}
\chi_S^{pion-Corr}=2\,{\langle\bar q q\rangle_{vac}^2\over f^4_\pi}
 \,{3\over 2}\,\int{id\omega d{\bf q}\over 
(2\pi)^4}\,\,{\bf q}^2\,\left[\right(D^2_\pi  D_{0\pi}\,+\,D_\pi  D^2_{0\pi}\left)
\,\tilde\Pi^0
-2\, D^3_{0\pi}\,\,\Pi^0\right](\omega,{\bf q}).\label{CHICORR}
\end{equation}
\\
In the evaluation of the nuclear contribution, $\chi_S^{nuclear}$,
the derivative $(\partial\bar S /\partial c)_\mu$ is  obtained by taking the derivative of the
minimization equation with respect to the parameter $c$~:
\begin{equation}
m^{*2}_\sigma \left({\partial\bar S \over\partial c}\right)_\mu=1\,-\,
g^*_S\, \,{M^*_N\over E^*_F}\,\Pi_0(0)\left[g^*_S  \left({\partial\bar S \over\partial c}\right)_\mu
\,+\,{E^*_F\over M^*_N}\,X^{(\pi)}\,+\,V_V\,\left({\partial \rho \over\partial c}\right)_\mu\right] ,
\label{PASDENOM}\end{equation}
with the auxiliary quantities~:
\begin{equation}
X^{(\pi)}={1\over 2 f_\pi }
\,{\partial \langle\Phi^2\rangle^{FC}\over\partial \rho}\,+\,
{M^*_N\over E^*_F}\,{\sigma_N^{(\pi)}\over f_\pi m_\pi^2},\qquad
V_V={E^*_F\over M^*_N}\,\left({g^2_\omega\over  m_\omega^2}\,+\,
{\partial^2\varepsilon^{FC}\over \partial \rho^2}\right)
\end{equation}
and with $\Pi_0(0)= - 2 M^*_N\, p_F/\pi^2$, which is  the 
non-relativistic free Fermi gas particle-hole polarization propagator 
in the Hartree scheme, at zero energy in the limit of vanishing momentum. The derivative of 
the baryonic  density is obtained by taking the derivative with respect to $c$ of 
eq. \ref{RHOB}, with the result~:
\begin{equation}
\left({\partial \rho \over\partial c}\right)_\mu=
\left(g^*_S\,\left({\partial\bar S \over\partial c}\right)_\mu\,+\,{E^*_F\over M^*_N}\,X^{(\pi)}\right)
\Pi_0(0)\,\big( 1\,-\,V_V\,\Pi_0(0)\big)^{-1}.
\end{equation}
It follows that  $\chi_S^{nuclear}$ can be written in the following form which displays
  the  propagator of the scalar field and the role of the pion loops in this propagation~:
\begin{equation}
\chi_S^{nuclear}=-2\,{\langle\bar q q\rangle_{vac}^2\over f_\pi^2}\,
\left({1\over  m^{*2}_\sigma}\,-\,{1\over  m^{*2}_\sigma}\,\left({\sigma_N^{(\pi)}\,+\,\sigma_N^{(\sigma)}
\over\sigma_N^{(\sigma)}}\right)^2_{eff}\Pi_{SS}(0)\,{1\over  m^{*2}_\sigma}\right)\,.
\label{CHISNUCL}
\end{equation}
Here $\Pi_{SS}(0)$  is the full scalar polarization propagators 
(in which we include the coupling constant)~:
\begin{equation}
\Pi_{SS}(0)=g^{*2}_S \,{M^*_N\over E^*_F}\,\Pi_0(0)\,
\left[1 -\,V_{res}\, \Pi_0(0)\right]^{-1}.\label{PISS}\,,
\end{equation}
where $V_{res}$ is the residual interaction~:
\begin{equation}
V_{res}={E^*_F\over M^*_N}\,\left({g^2_\omega\over  m_\omega^2}\,+\,
{ \partial^2\varepsilon^{FC}\over\partial \rho^2}\right)\,-\,
{M^*_N\over E^*_F}\,{g^{*2}_S\over m^{*2}_\sigma}\,.\label{INTRES}
\end{equation}
This expression differs from the mean-field one of our previous work \cite{CE05} by the presence 
of the term in $\partial^2\varepsilon^{FC}/\partial \rho^2.$
In the expression \ref{CHISNUCL} we have introduced quantities that we denote effective sigma commutators  
with the explicit  expressions  ~:
\begin{equation}
(\sigma_N)_{eff}=(\sigma_N^{(\pi)})_{eff}\, + \,(\sigma_N^{(\sigma)})_{eff}=
{M^*_N\over E^*_F}\,\sigma_N^{(\pi)}\, +\, {m_\pi^2\over 2} 
{\partial \langle\Phi^2\rangle^{FC}\over\partial \rho}\,+\,
{M^*_N\over E^*_F}\,f_\pi\, g^*_S\, {m^2_\pi\over m^{*2}_\sigma}.\label{SIGMAEFF}
\end{equation}
It is interesting to compare it with the effective sigma commutator governing the
 evolution of the
chiral condensate.which is~:
\begin{equation}
(\tilde\sigma_N)_{tot}=(\tilde\sigma_N^{(\pi)})\, + \,(\tilde\sigma_N^{(\pi)})=
{\rho_S\over\rho}\,\sigma_N^{(\pi)}\, +\, {m_\pi^2\over 2} 
{\langle\Phi^2\rangle^{FC}\over\rho}\,+\,
f_\pi\, m^2_\pi\,{\bar s\over \rho}.\label{SIGMATOT}
\end{equation}
As was shown in ref. \cite{CE03}, this last quantity, corresponding to the full
nuclear sigma commutator per nucleon, also governs the evolution of the pseudoscalar
susceptibility according to~:
\begin{equation}
\chi_{PS}=-2\,{\langle\bar q q\rangle_{vac}^2\over f_\pi^2\,m^2_\pi}\,
\left( 1\,-\,{(\tilde\sigma_N)_{tot}\,\rho\over f_\pi^2\,m^2_\pi}\right)\,.\label{CHIPS}\
\end{equation}
Comparing the expressions \ref{SIGMAEFF} and  \ref{SIGMATOT} we see that they are not identical. They 
coincide to leading order in density but the higher order terms differ. In particular  
$\langle\Phi^2\rangle^{FC}/\rho$ is replaced in eq. \ref{SIGMAEFF}
 by $\partial\langle\Phi^2\rangle^{FC}/\partial \rho$.
We recover in this work the result of \cite{CEG03}, that the  quantity which
governs the transformation of the quark scalar density fluctuations into nucleonic ones
represents an effective  nucleon sigma commutator
which includes pion loops.
In  ref. \cite{CEG03} it was the free nucleon one while here interactions are incorporated
and we deal with effective values. Our present result thus generalizes the first order result
of \cite{CEG03}. 

Coming back to the expression  \ref{CHISNUCL} of $\chi_S^{nuclear}$, we point out that the
pion loops only enter in the effective sigma commutator which includes medium effects.  
The mass, $m^*_\sigma$, which enters the
denominators  is instead unaffected by the pion loops. The nuclear
 contribution  $\chi_S^{nuclear}$ is
indeed linked to the propagation of the chiral invariant scalar field $s$ which is not
coupled to pions. The pure pion cloud contribution (eq. \ref{CHIPI}) is totally decoupled. 
This point was already discussed, within a different approach in ref. \cite{CDEM06}.

\section{Numerical results and discussion}
\subsection{Fixing the parameters}
One important constraint to fix the parameters is the pion-nucleon sigma term which has a value,  
$\sigma_N\simeq 50 \, MeV$. In our approach it receives two contributions: one from the pion cloud, 
the other from the sigma meson. We point out that if our model represents 
a bosonized  underlying NJL model, the latter contribution represents the sigma term of
constituent quarks. It is interesting to make a  connection with the lattice evaluations 
of the evolution of the nucleon mass with the pion mass.
Indeed lattice simulations of the nucleon mass as a function of the squared pion mass 
(equivalently  the quark mass) are available in the mass region  beyond $m_{\pi } \simeq 400MeV$. 
The derivative ${\partial M_N/ \partial m^2_{\pi}}\, =\sigma_N/m_{\pi}^2$ provides the nucleon sigma 
commutator.  In turn the derivative of $ \sigma_N$ leads to the susceptibility. Both  quantities are  
strongly influenced by the pion cloud  which has a non-analytic behavior in the quark mass, preventing 
a polynomial expansion in this quantity. However, it is possible to extrapolate lattice data 
using  chiral models of the nucleon as discussed  by Thomas and collaborators and we use their 
recent version ref.  \cite{TGLY04}. In their work the pionic self-energy  contribution to 
the nucleon mass is separated out using different cut-off forms for the pion loops 
(gaussian, dipole, monopole) with an adjustable parameter $\Lambda$. 
They expanded the remaining part  in terms of $m^2_{\pi}$  as follows~:
\begin {equation}
M_N(m^{2}_{\pi}) = 
a_{0}\,+\,a_{2}\,m^{2}_{\pi}\, +\,a_{4}\,m^{4}_{\pi}\,+\,\Sigma_{\pi}(m_{\pi}, \Lambda)\,.
\end{equation}
The best fit value of the parameter $a_{4}$  which fixes the 
susceptibility shows little sensitivity to the shape of the form 
factor, with a value 
$a_4 \simeq- 0.5\, GeV^{-3} $ while $a_2 \simeq 1.5\,GeV^{-1}$ (see ref. \cite{TGLY04})),
from which we can infer the non-pionic pieces of the sigma commutator~: 
\begin{equation}
 \sigma_N^{non-pion} = m^2_{\pi} {\partial M\over \partial m^2_{\pi }}
=a_2 \,m^2_{\pi}\, + \,2\,a_4 \, m^4_{\pi}\simeq 29\, MeV\,.
\end{equation}
This  number indicates the existence of a large component  $\sigma_N$ beside the pion one, that it 
is natural to attribute to the scalar field.   
The identification of   $\sigma_N^{non-pion}$ with $\sigma_N^{(\sigma)}$ of our model
fixes the sigma mass to a value  $m_\sigma=800\,MeV$, close to the one $\simeq750 MeV$ that we have used 
in our previous article \cite{CE05}. As it is the ratio $g_s/{m^2_\sigma}$ which is thus determined this 
value of $m_\sigma$ is associated with the coupling constant of the linear sigma model 
$g_S=M_N/f_\pi=10$. Lowering $g_S$ reduces $m_\sigma$. With 
our  cutoff , $\Lambda=0.98\,GeV$, which  yields $\sigma_N^{(\pi)}=21.5 MeV$,  the 
total value of the sigma term is $\sigma_N=50.5 \, MeV$, a quite satisfactory result.

Although the procedule  becomes 
more uncertain, on can also try to extract the non-pionic part of the nucleon scalar susceptibility~:
\begin{equation}
 \chi_{NS}^{non- pion}~= 2{\langle\bar q q\rangle_{vac}^2 \over f^4_{\pi} }
 {\partial~~\over \partial m^2_{\pi }}\left({\sigma_N^{non-pion} \over m^2_{\pi }}\right)=  
 {\langle\bar q q\rangle_{vac}^2 \over f^4_{\pi} }\,4\,a_{4}\,.
\end{equation}
It has to be compared with the non-pionic piece of the nucleon scalar susceptibility. This can be
done taking the zero density limit of eq. \ref{CHISNUCL}~:
\begin{equation}
 \chi_{NS}^{(\sigma)}= -2{\langle\bar q q\rangle_{vac}^2 \over f^2_{\pi} }\,
 \left({1\over m^{*2}_\sigma}\,-\,{1\over m^{2}_\sigma}\right)\,{1\over\rho}=
-2{\langle\bar q q\rangle_{vac}^2 \over f^2_{\pi} }\,{1\over m^{4}_\sigma}\,
\left({3\,g_S\over f_\pi}\,-\,\kappa_{NS}\right)\,. 
\end{equation}
This identification allows an estimate of the crucial parameter $\kappa_{NS}$.
For convenience, as in \cite{CE05}, we introduce the dimensionless parameter
$C =(f^2_\pi /2 M_N)\kappa_{NS}$ and we obtain~:
\begin{equation}
a_4=-{f_\pi\,g_S\over 2\,m^{4}_\sigma}\,(3\,-\,2\,C)\,.
\end{equation}   
The value $a_4 \simeq- 0.5\, GeV^{-3} $ leads to $C \simeq 1.25$, which indicates a strong cancellation 
between the two components of $a_4$, which  makes the results sensitive to the exact values of the 
parameters. 
Therefore, due to the uncertainties of the procedure, we take this value of $C$ only as indicative.

Hence for the following  discussions we fix the various parameters  
according to $m_\sigma=800\,MeV$, $\Lambda=980\,MeV$, 
$g'_{NN}=0.7,\,g'_{N\Delta}=0.3,\, g'_{\Delta\Delta}=0.5$, $C_\rho=2$ and the parameter associated 
with the nucleonic response is allowed to vary in a broad region around $C = 1.25$. The vector 
coupling constant $g_\omega$ is a totally free parameter. Finally, as in ref. \cite{CE05}, we also 
impose that the  nucleonic response vanishes at the chiral restoration ($\bar s=-f_\pi$)~: 
\begin{equation}
\kappa_{NS}(\bar s)={\partial^2M_N\over\partial \bar s^2}=
\kappa_{NS}\left(1\,+\,{\bar s\over  f_\pi}\right).
\end{equation}
Accordingly,  the effective nucleon mass of eq. \ref{EFFMASSN} is modified according to~:
\begin{equation}
M^*_N(\bar s)=M_N\,\left( 1+{\bar s\over f_{\pi }}\right)\,+{1\over 2}\,\kappa_{NS}\,
\bar s^2\,\left(1\,+\,{\bar s\over 3\,f_\pi}\right) 
\end{equation}
and  the effective scalar coupling constant becomes dependent on $\bar s$:
\begin{equation}
g^*_S(\bar s)={\partial M^*_N\over\partial\bar s}={M_N\over f_{\pi}}
\,+\,\kappa_{NS}\, \bar s\,\left(1\,+\,{\bar s\over 2\, f_\pi} \right).
\end{equation}
 \begin{figure}                                        
\centering
\includegraphics[width=0.5\linewidth,angle=270]{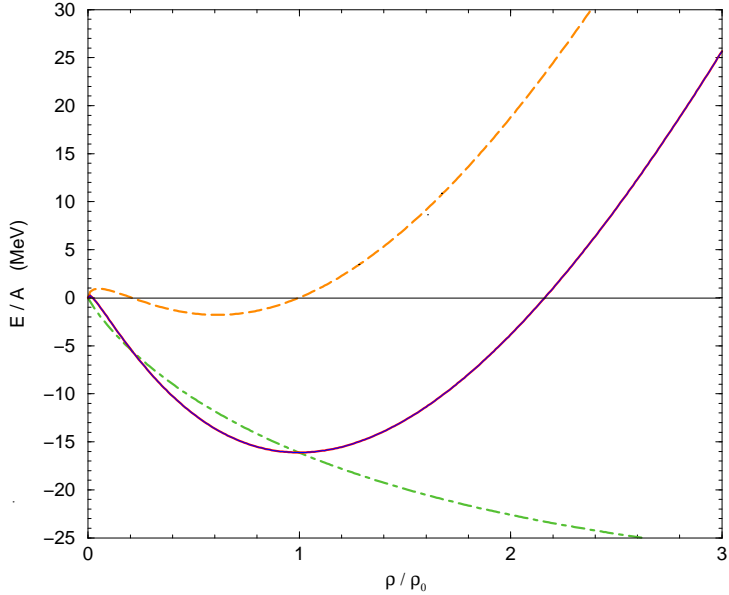} 
\caption{Binding energy of nuclear matter with $g_\omega=7.3$ and $C=0.985$ 
keeping only the Fock term on top of $\sigma$ and
$\omega$ exchange.
The full line corresponds to the full result, the dotted line represents the binding energy
without the Fock term and the dot-dashed line corresponds to the contribution of the Fock term.}
\label{fockbind}
\end{figure} 
\begin{figure}                      
\centering
\includegraphics[width=0.5\linewidth,angle=270]{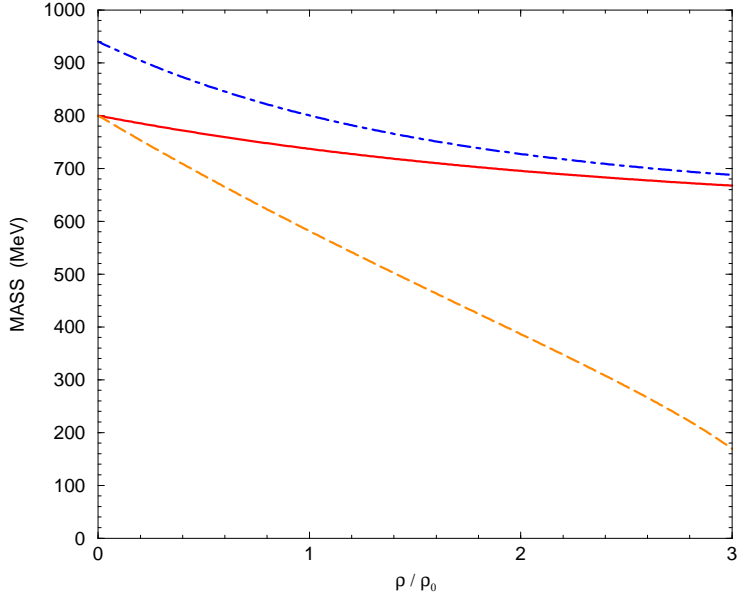} 
\caption{Density evolution of the nucleon (dot-dashed curve)
and sigma masses (full curve) with $g_\omega=7.3$ and $C=0.985$; the dashed
curve corresponds to the sigma mass  when  the effect of the nucleon susceptibility is removed.}
\label{fockmass}
\end{figure} 
\begin{figure}                    
\centering
\includegraphics[width=0.5\linewidth,angle=270]{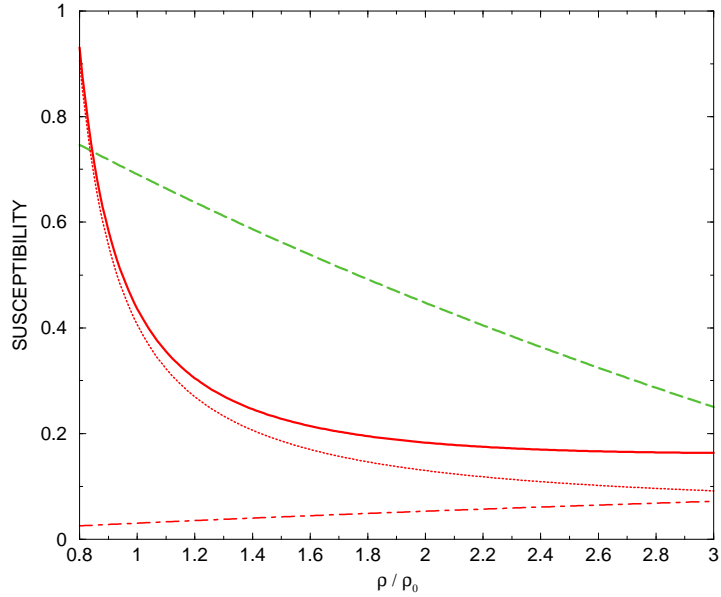} 
\caption{Density evolution of the QCD susceptibilities (normalized to the
  vacuum value of the pseudoscalar one)  with $g_\omega=7.3$ and $C=0.985$ 
keeping only the Fock term on top of $\sigma$ and
$\omega$ exchange.
Dashed curve: pseudoscalar susceptibility. Full curve: scalar susceptibility. Dotted curve:
nuclear contribution to the scalar susceptibility. Dot-dashed curve: pion loop contribution to the
scalar susceptibility.}
\label{focksusc}
\end{figure} 
\begin{figure}                    
\centering
\includegraphics[width=0.5\linewidth,angle=270]{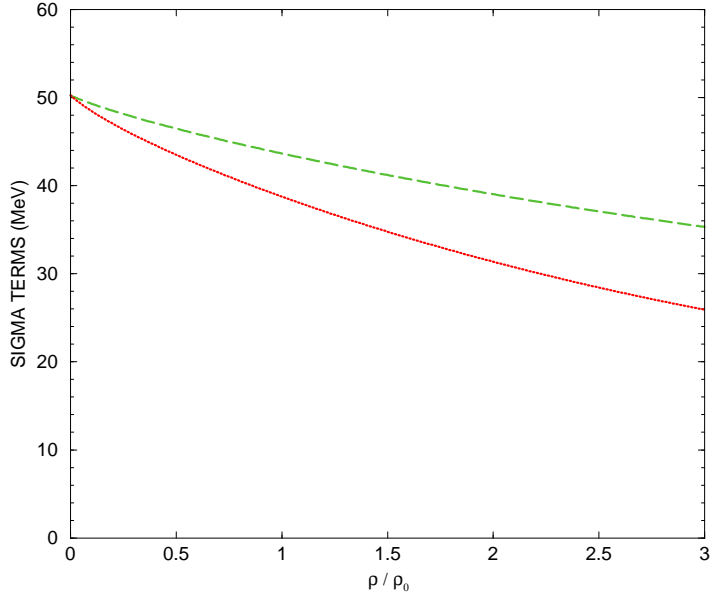} 
\caption{Density evolution of the full in-medium sigma term, $(\tilde\sigma_N)_{tot}$ (dotted line) 
and of the effective sigma term, $(\sigma_N)_{eff}$ (full line), with $g_\omega=7.3$ and $C=0.985$ 
keeping only the Fock term on top of $\sigma$ and
$\omega$ exchange. }
\label{focksigma}
\end{figure}
\begin{figure}                    
\centering
\includegraphics[width=0.5\linewidth,angle=270]{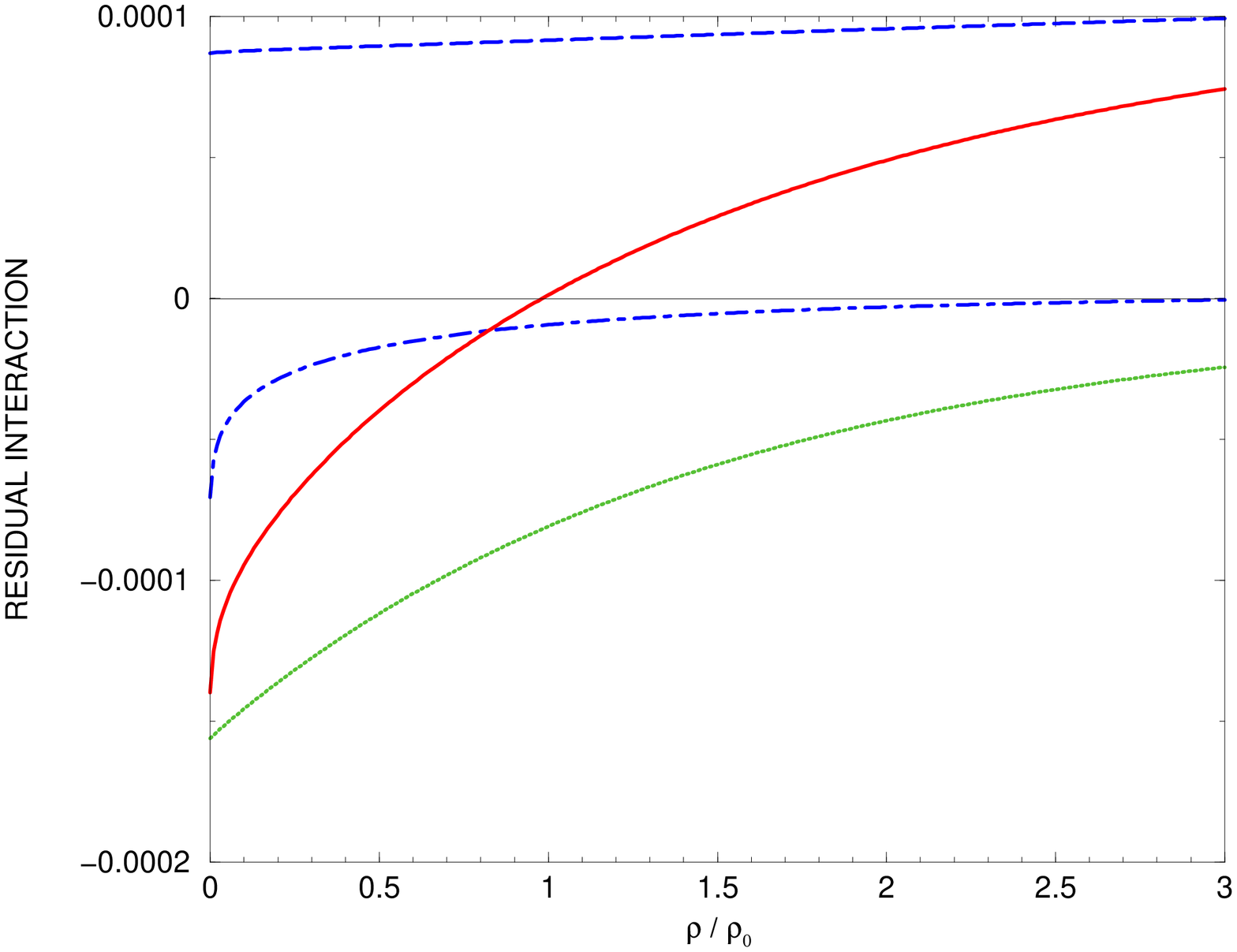} 
\caption{Density evolution of the residual interaction in $MeV^{-2}$ (continuous line). 
Dashed line: omega exchange contribution.
Dot-dashed line: Fock term contribution. Dotted line: sigma exchange contribution.
   }
\label{fockrespot}
\end{figure} 

\subsection{Results with the influence of the Fock term}
The result of the calculation for the binding energy of nuclear matter in presence of the 
Fock term only ({\it i.e.}, ignoring 
the correlation energy) is displayed in fig. \ref{fockbind}. 
The saturation point($\rho_0=0.16\,fm^{-3}$, $E/A=- 
16.1\,MeV$) is  obtained with $C=0.985$ and $g_\omega=7.3$. The corresponding incompressibility 
is $K=254 \,MeV$. Comparing with our previous results \cite{CE05} we see that the
 inclusion of the Fock term somewhat 
 improves the description since it lowers $K$. We observe in fig. \ref{fockbind} that
the value of the Fock term at saturation density is nearly the same as the binding energy. 
It turns out that, at this  density, this binding comes 
from the transverse Fock term ($g'$ + rho exchange). In the longitudinal channel there is an almost 
exact compensation between the pion  and the $g'$ exchanges.  
The genuine Fock term associated 
with the Yukawa piece of the pion exchange represents   an attractive contribution to $E/A$ of only 
$-4.2 MeV$ $\simeq 20\%$ smaller than what is obtained in 
relativistic theories \cite{BMVN87} where the $\pi NN$ form factor is omitted.

For completeness we also display in this subsection (fig. \ref{fockmass}) 
the nucleon and sigma masses with these 
values of the parameters, although they are not sensitive to the inclusion of the Fock term. 
At normal nuclear matter density the effective nucleon mass is close 
to $800\, MeV$. As already emphasized in  our previous  work \cite{CE05} the sigma mass remains instead 
quite stable, due to the introduction of the nucleonic scalar response.

The evolution of the susceptibilities is depicted in fig. \ref{focksusc}. The convergence 
effect between the pseudoscalar susceptibility (dashed line) and the scalar one (full curve) is  
more pronounced than in the mean field approach of our previous work 
\cite{CE05}. The reason is clear and linked to the  introduction of pion loops. Firstly the evolution 
of the pseudoscalar susceptibility, $\chi_{PS}$, as written in eq. \ref{CHIPS}, is governed by the 
in-medium sigma term,  $(\tilde\sigma_N)_{tot}$, which in the present approach
receives a contribution from the pion loops  which increases its value
(see eq. \ref{SIGMATOT}). Secondly, for the scalar susceptibility, the effective sigma term, 
$(\sigma_N)_{eff}$, transforming   the scalar quark density fluctuations into nuclear excitations also 
has a pion cloud contribution  (eq. \ref{SIGMAEFF}) and this effect
increases the nuclear part, $\chi_S^{nuclear}$  (dotted curve), of the scalar susceptibility. 
Finally the scalar susceptibility also contains a pure pion loop contribution, 
$\chi_S^{pion loop}$  (dot-dashed curve), which becomes more important  at large density.
For illustration we also compare in fig. \ref{focksigma}
the in-medium modified nucleon sigma term $(\tilde\sigma_N)_{tot}$ 
governing $\chi_{PS}$ (dotted curve) and the effective  sigma term $(\sigma_N)_{eff}$ (full curve) 
affecting the nuclear piece $\chi_S^{nuclear}$. 

The density behavior of the residual interaction (eq. \ref{INTRES}) is displayed in fig. \ref{fockrespot}
with  separate curves for the various components  (omega and sigma exchanges, Fock contribution). 
While the $\omega$
exchange has a very smooth density dependence, one notices the rapid variation of the sigma component
associated with the dropping of the effective scalar coupling constant $g^*_S$ with increasing density.
Around $\rho_0$ $V_{res}$ turns from attraction into repulsion. 
This feature comes in part from the Fock term evolution and mostly
from the behavior of the sigma exchange contribution  
\begin{figure}                     
\centering
\includegraphics[width=0.5\linewidth,angle=0]{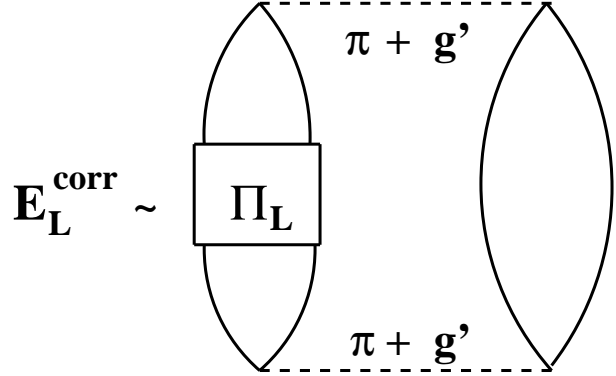} 
\caption{Schematic representation of the longitudinal spin-isospin contribution 
to the correlation energy.}
\label{encorr}
\end{figure} 
\begin{figure}                     
\centering
\includegraphics[width=0.5\linewidth,angle=270]{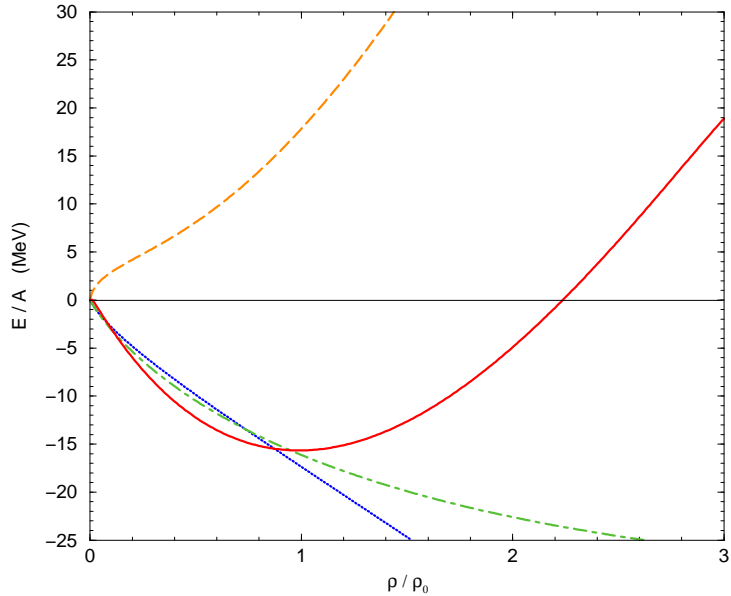} 
\caption{ Binding energy of nuclear matter with $g_\omega=8$, $m_\sigma =850\,MeV$ and $C=0.985$ 
with the Fock  and correlation energies on top of $\sigma$ and
$\omega$ exchange.
The full line corresponds to the full result, the dotted line represents the binding energy
without the Fock and  correlation energies and the dot-dashed line corresponds to 
the contribution of the Fock terms. The decreasing dotted line (always negative) represents the
correlation energy.}
\label{corrbind}
\end{figure} 
\begin{figure}                     
\centering
\includegraphics[width=0.5\linewidth,angle=270]{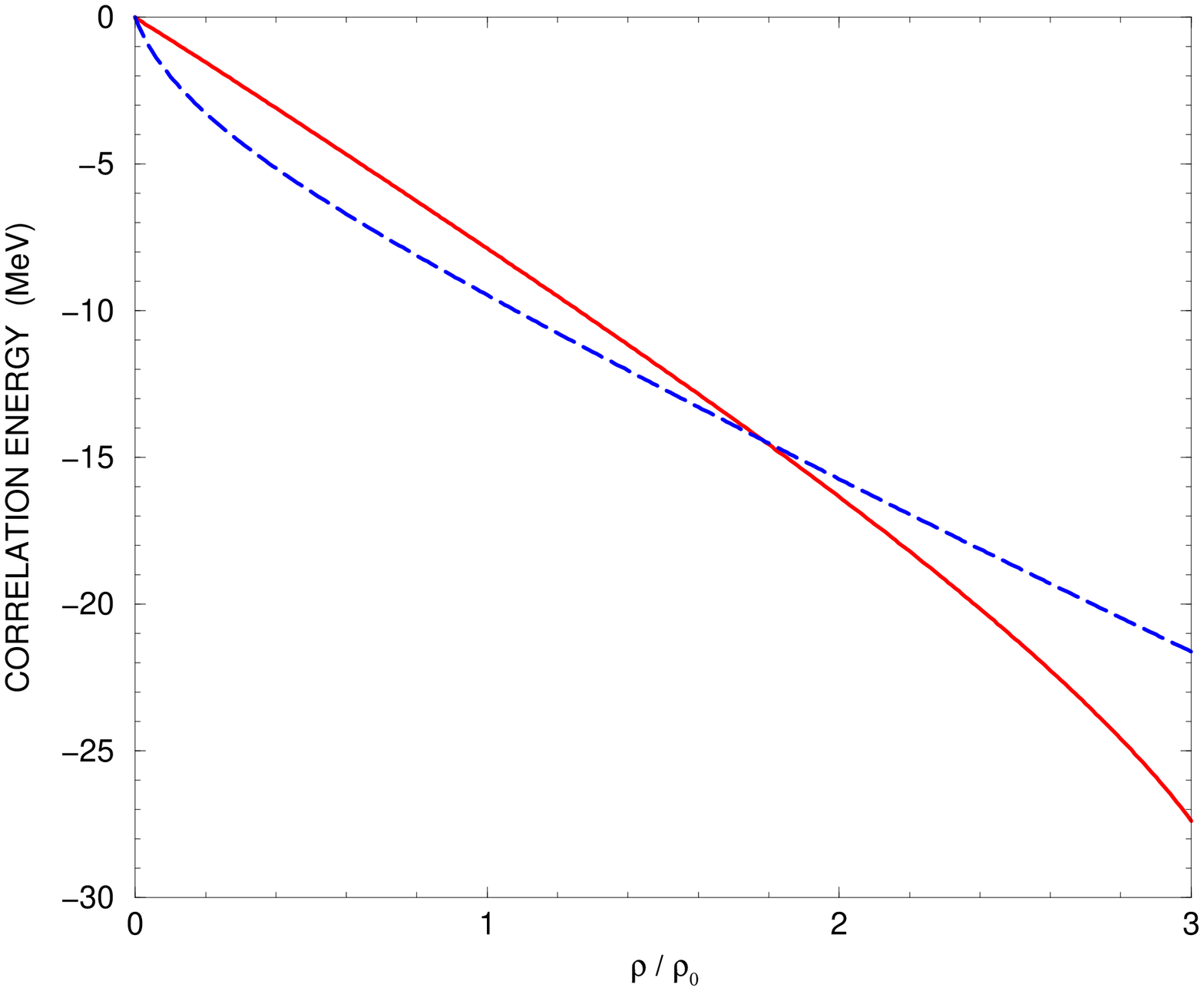} 
\caption{ Density evolution of the longitudinal (full line) and transverse (dashed line)
contributions to the correlation energy for $g'_{N\Delta}=0.3$.}
\label{corrlt03}
\end{figure} 
\begin{figure}                     
\centering
\includegraphics[width=0.5\linewidth,angle=270]{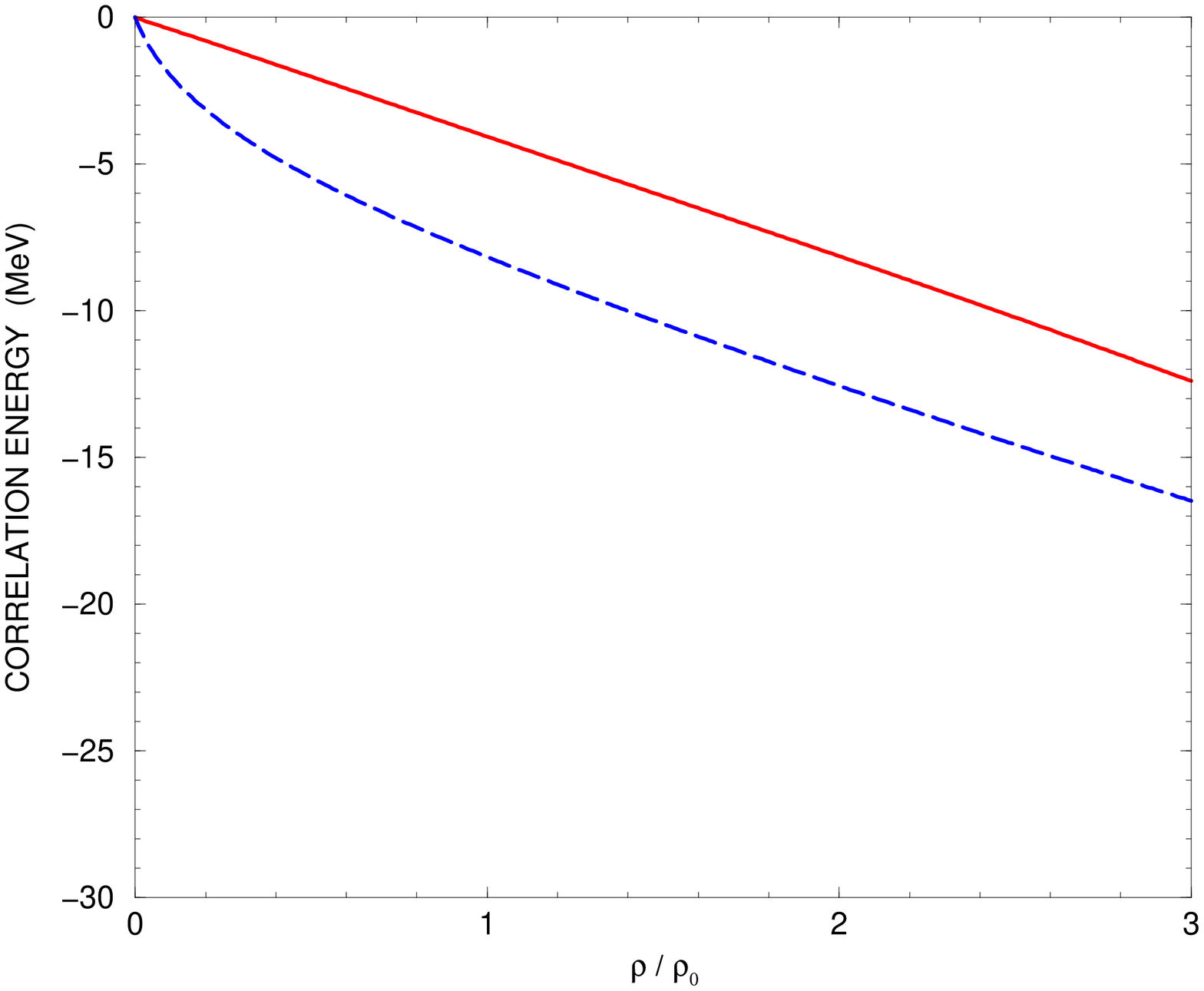} 
\caption{Density evolution of the longitudinal (full line) and transverse (dashed line)
contributions to the correlation energy for $g'_{N\Delta}=0.5$.  }
\label{corrlt05}
\end{figure} 
 \begin{figure}                  
\centering
\includegraphics[width=0.5\linewidth,angle=270]{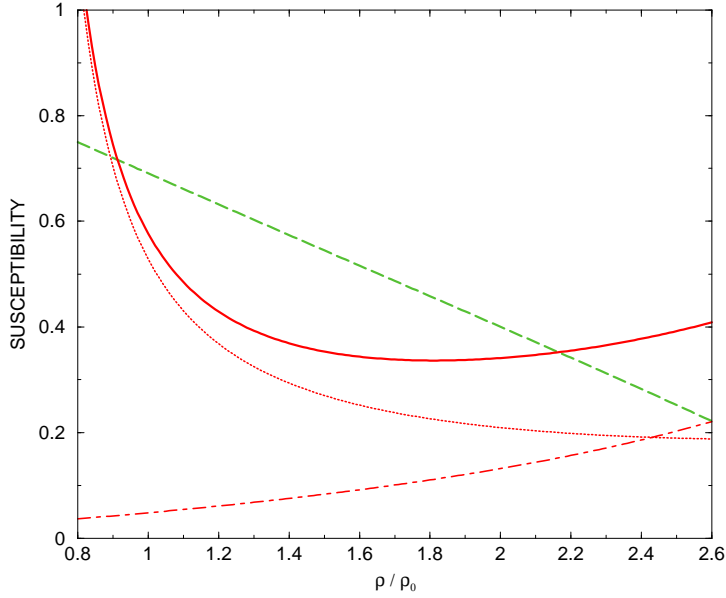} 
\caption {Density evolution of the QCD susceptibilities (normalized to the
  vacuum value of the pseudoscalar one)  with $g_\omega=8$, $m_\sigma=850\, MeV$ and $C=0.985$ 
with the Fock  and correlation terms on top of $\sigma$ and
$\omega$ exchange.
Dashed curve: pseudoscalar susceptibility. Full curve: scalar susceptibility. Dotted curve:
nuclear contribution to the scalar susceptibility. Dot-dashed curve: pion loop contribution to the
scalar susceptibility.}
\label{corrsusc}
\end{figure} 
\begin{figure}                  
\centering
\includegraphics[width=0.5\linewidth,angle=270]{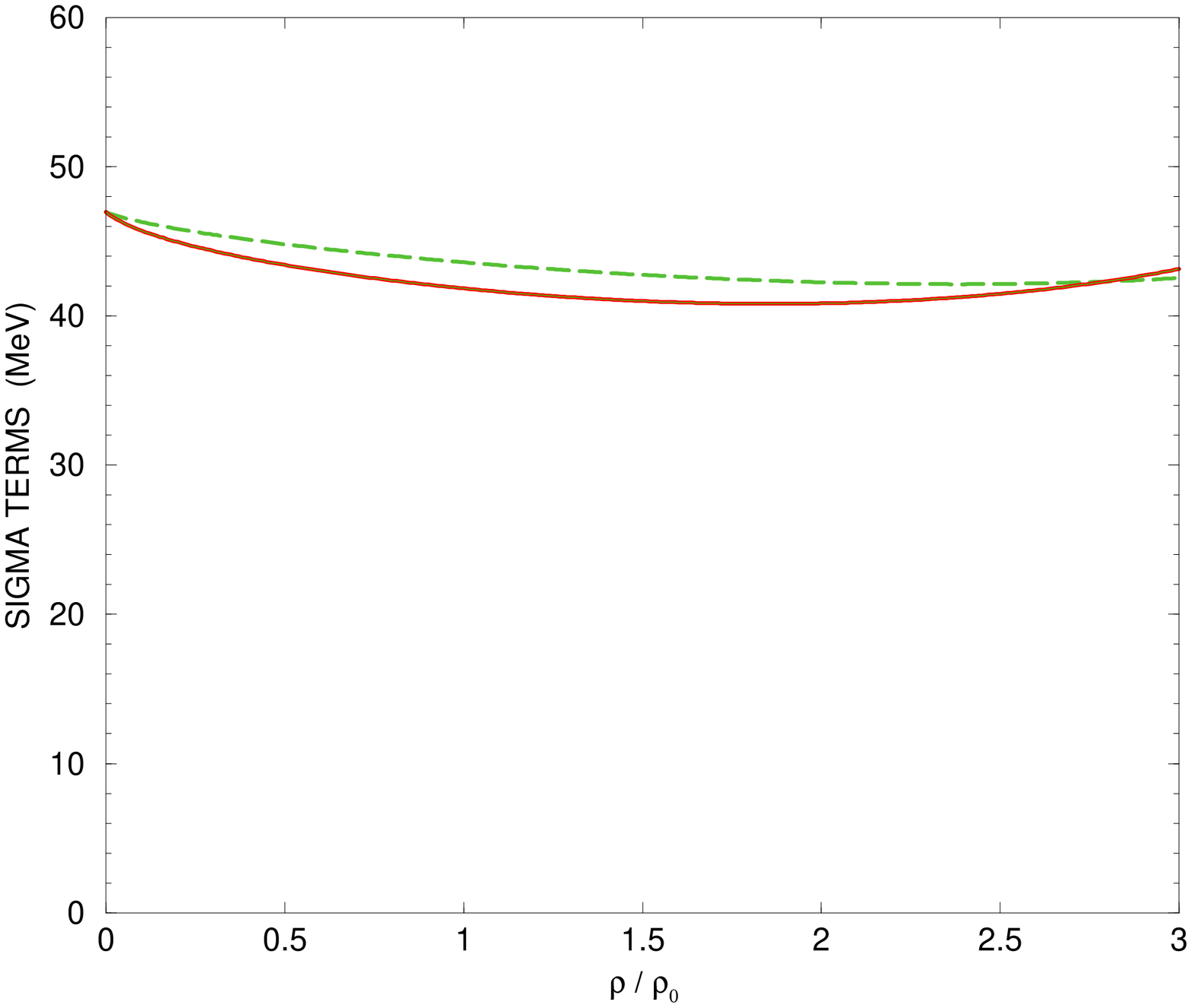} 
\caption{Density evolution of the full in medium sigma term, $(\tilde\sigma_N)_{tot}$ (dotted line) 
and of the effective sigma term, $(\sigma_N)_{eff}$ (full line),with $g_\omega=8$, $m_\sigma=850\, MeV$ 
and $C=0.985$  with the Fock  and correlation terms on top of $\sigma$ and
$\omega$ exchange.}
\label{corrsigma}
\end{figure}

\subsection{Influence of the correlation term}
The expression of the longitudinal contribution to the correlation energy, which is always attractive, 
has  been given in eq. \ref{EL}. It is  is depicted schematically in fig. \ref{encorr} 
and involves the full longitudinal spin-isospin polarization propagator $\Pi_L$, which is calculated 
in practice within the RPA ring approximation. There is a similar expression for the transverse
piece.
 
With the previous choice of these parameters it turns out 
that the numerical value of the correlation energy is close to the Fock term energy. 
This extra attraction has to be compensated by an additionnal repulsion in order to account
for  the saturation point.  Increasing only the vector coupling constant
tends to give too low a saturation density. 
However a slight adjustment of the sigma mass yields a correct saturation point. Taking   
$g_\omega=8$ and $m_\sigma=850\, MeV$ (corresponding to $ \sigma_N^{non-pion}=26\, MeV$) 
and keeping $C=0.985$  we obtain the saturation point~: 
$\rho_0=0.16\,fm^{-3}$, $E/A=- 15.7\,MeV$. 
The corresponding correlation energy is  $E^{Corr}/A=-17.4\, MeV$ 
which divides into a longitudinal  piece, $E^{Corr}_L/A=-7.9\, MeV$ and a transverse one, 
$E^{Corr}_T/A=-9.5\, MeV$.  The results of the calculation are shown in fig. \ref{corrbind}.
We stress the importance of the transverse channel in the correlation energy as compared to the 
longitudinal one. The reason is the strong screening of the pion exchange by the
short-range interaction in the longitudinal channel, hence a sensitivity to the $g'$ parameters. 
For instance changing  $g'_{N\Delta}$ from $0.3$ to $0.5$ reduces further the longitudinal value to 
$E^{Corr}_L/A=-3.7\,MeV$ while the transverse contribution shows
less sensitivity, with $E^{Corr}_T/A=-7.8\,MeV$, such that the total value becomes
 $E^{Corr}/A=-11.5\, MeV$. With the value  $g'_{N\Delta}=0.3$
the longitudinal contribution to the correlation energy is dominated by $2\pi$ exchange with one Delta 
intermediate excitation. Iterated  pion exchange is smaller, because of the screening effect. 
Therefore,  as the main mechanism with one Delta is a part of the $NN$ potential, 
we expect the longitudinal correlation energy  to be linear in density,
 at variance with the transverse channel where the Delta 
is not dominant. These behaviors are illustrated in fig. \ref{corrlt03}. The longitudinal
correlation energy has indeed a linear behavior at low densities. It deviates from 
linearity at high density where it becomes more attractive. This reflects  the appearance of 
many body forces due to the increase of the nuclear pionic field, the critical opalescence effect 
\cite{DE78} induced by the small value of $g'_{N\Delta}$. Taking for illustration 
a larger g' value,  $g'_{N\Delta}=0.5$, the linearity is fulfilled in a larger density range, as 
shown in fig. \ref{corrlt05}.  

It  is interesting to make a comparison with the chiral perturbation
calculation of ref. \cite{KFW02} where the effect of short-range correlation is absent. 
In the latter work the contribution of the iterated pion exchange is 
found to be $-68\, MeV$,  the  corresponding Hartree diagramm coinciding with the lowest order term
of the longitudinal correlation energy depicted on fig. \ref{encorr}. There  
is nearly one order of magnitude difference between the two results, owing to the fact that
in our phenomenological approach pion exchange is strongly screened by short-range correlations.

Although in detail our numerical results on the correlation energy present a  sensitivity to various 
factors,  the values of  the $g'$ parameters, the rho coupling ($C_\rho$)  or the form factors,  
especially in the transverse channel, we believe that our conclusion about the  moderate importance
of the correlation part of the energy is robust. The important 
point is that  the rest of the interaction, 
as  provided by lattice data, is sufficient to bind the nuclear system.

In the spirit our approach has a similarity with the density functional approach of Finelli 
{\it et al} \cite{FKVW06}. As them we have  background scalar and vector fields. The first one
has a connection with chiral symmetry restoration of QCD. However there is an essential 
difference concerning the scalar field. For Finelli {\it et al.}  the full nucleon 
sigma commutator enters the scalar self-energy of the nucleon according to~:
\begin{equation}
\Sigma^{(0)}_S=-{\sigma_N\,M_N\over f_\pi^2\ m^2_\pi}\,\rho_S .
\end{equation}
In our case instead, only part of the nucleon sigma commutator enters this self-energy.
It is the part which originates from the chiral invariant field $s$. To 
leading order in density we have~:
\begin{equation}
\Sigma^{(0)}_S=M_N\,{\bar s\over f_\pi}= -{\sigma_N^{(\sigma)}\,M_N\over f_\pi^2\,m^2_\pi}\,\rho_S.
\end{equation}
The other component from the nucleon pion cloud, $\sigma_N^{(\pi)}\simeq 21\,MeV$ should not enter the 
mass evolution as it is cancelled by other terms, as imposed by chiral constraints. This point  
was emphasized in several works, see {\it e.g.} reference \cite{B94,CDEM06}. 
Accordingly the link between our nuclear scalar field and QCD does not occur through
the total nucleon sigma commutator but only through part of it, its non-pionic piece.
Information on this last quantity can be obtained from the lattice result on the nucleon evolution with
pion mass, once the pionic contribution is separated out, as done in ref. \cite{TGLY04} and   explained
in section {\bf 4}. Our result is therefore not totally  model independent but this limitation is imposed 
by chiral constraints \cite{B96}.

For the evaluation of the scalar susceptibility including correlation terms, we make the
assumption that the pionic correlation energy density and the correlated part of the pion scalar
density go both like $\rho^2$, a reasonable approximation in view of the results shown on 
fig. \ref{corrlt03}. The net conclusion is that the
convergence effect between the scalar and pseudoscalar susceptibilities is much more pronounced
as shown in  fig. \ref{corrsusc} (compare with fig. \ref{focksusc}). 
One reason is that  the pionic piece of the scalar susceptibility, 
$\chi_S^{pion loop}$, is significantly increased by the RPA correlations, a conclusion already reached
in our previous work \cite{CDEM06}. Another reason comes from the fact that the full in-medium sigma
term $(\tilde\sigma_N)_{tot}$, governing the pseudoscalar susceptibility and the effective 
sigma term,  $(\sigma_N)_{eff}$ affecting the nuclear piece of the scalar susceptibility remain
relatively stable with the density, as depicted on fig. \ref{corrsigma}, 
while they significantly decreased in fig.\ref{focksigma}.

\section{Conclusion}
We have studied in this work a relativistic nuclear model based on a chiral version of the
$\sigma$ and $\omega$ exchange model. We have worked beyond the mean field approximation
introducing the pion loops. Our goal is  to reach a consistent description of
matter which can apply at densities larger than the saturation one 
by introducing the pion loops on top of a mean field approach. The aim is to reach the 
best possible description in the framework of  the model.The pion is an important actor
of the nuclear dynamics and a credible theory should incorporate its effect.
 
In establishing the model and fixing its parameters we have kept contact with other domains. 
In particular we introduce beside the pion 
the short-range components, which are indissociable from the pion in the spin-isospin residual
interaction. They are embedded in the Landau-Migdal parameters and for their values
we have used the most recent informations. They favor a clear deviation from
universality, with a small value of $g'_{N\Delta}\simeq 0.3$. We have also insured
the compatibility of our parameters with informations from QCD. One important constraint is 
the total nucleon sigma commutator. But more information is also available from the lattice results 
on the evolution of the nucleon mass with the quark mass. From the analysis of these data by 
Thomas {\it et al.} \cite{TGLY04} we extracted the non-pionic piece of the sigma commutator. In our model this
is the scalar field contribution and it fixes the sigma mass to be used in our inputs. Another piece of 
information  is obtained from the higher term of the expansion, even if we take this information as 
indicative.  
Combining the informations from spin-isospin  physics and from QCD lattice
results we can reach a tenable description of the saturation properties. 
>From spin-isospin physics, we have taken the values of the Landau-Migdal $g'$ parameters to be used in 
association with pion and rho exchanges. They limit the value of the correlation energy 
from the  pion loops through a suppression of pion exchange by the short-range component. 
On the other hand the value 
of the non-pionic piece of the sigma commutator from lattice QCD, which fixes the sigma mass,
provides the  attraction needed for saturation to occur.
Moreover the higher term $a_4$ of the nucleon mass expansion 
strongly limits  the many-body effects that can occur in the propagation of the nuclear scalar
field. Otherwise the tadpole terms chiral theories alone would induce too much attraction and 
destroy saturation. 
It is quite remarkable than the parameters which reproduces  the saturation properties are compatible 
to the lattice ones. Nowhere have we faced a contradiction. It was not {\it a priori} obvious that
the amount of attraction needed from the scalar field fits the value of the 
non-pionic sigma commutator from lattice data. Nor that the amount of cancellation necessary in the 
medium effects of the sigma propagation from the nucleon scalar response is reproduced in the expansion 
of the lattice results, with values of the parameter $C$ remarkably close ($C=1.25$ for the lattice results 
versus $C\simeq 1$ in our fit). This consistency gives credibility to the description..

In more details, we have investigated the role  of the Fock term and of the correlations. 
For the second one we have 
found a moderate value, $\simeq - 17\, MeV$ at $\rho_0$ due to a suppression of the pion 
contribution by the short-range component in such a way that the transverse channel linked to 
rho exchange becomes dominant. The longidudinal component of the correlation energy 
is linear in density but it evolves faster at high densities, signalling the possible appearance of
many-body forces of attractive nature. They arise from the enhancement of the
nuclear pion field, the critical opalescence effect which  softens the equation of state at
large densities.
The compatibility of the parameters of our approach  with the QCD lattice data
is a support for our results on the density evolution
of the QCD quantities, the quark condensate and the QCD suceptibilities.We have insured that these 
results are in addition 
fully compatible with the saturation properties of nuclear matter.  For the quark condensate we find 
small deviations from a linear behavior with density, in spite of the inclusion of various types 
of many-body effects. This  also applies for the pseudoscalar
susceptibility which follows the condensate. The scalar susceptibility is instead
very sensitive to the interactions and shows a large enhancement as compared 
to the free value. It even surpasses the pseudoscalar one beyond $\rho \simeq
2.2\, \rho_0$ while in the vacuum it is much smaller.

Our description could be improved in various directions such as the incorporation of all the Fock terms 
(sigma and omega exchange, time component of the rho meson exchange) and a covariant description of the
short-range interaction in a fully covariant framework. It would also be interesting to   
extend the approach to the case of asymmetric nuclear  matter and neutron matter.

\section*{Acknowledgments} 
We thank P. Guichon for fruitful discussions.


\end{document}